\documentclass[sigconf,authorversion]{acmart}

\copyrightyear{2026}
\acmYear{2026}
\setcopyright{cc}
\setcctype{by}
\acmConference[CHI '26]{Proceedings of the 2026 CHI Conference on Human Factors in Computing Systems}{April 13--17, 2026}{Barcelona, Spain}
\acmBooktitle{Proceedings of the 2026 CHI Conference on Human Factors in Computing Systems (CHI '26), April 13--17, 2026, Barcelona, Spain}
\acmDOI{10.1145/3772318.3794641}
\acmISBN{979-8-4007-2278-3/2026/04}

\acmSubmissionID{2181}

\begin{document}

\title[``Chasing Shadows'': Understanding Personal Data Externalization and Self-Tracking for Neurodivergent Individuals]{``Chasing Shadows'': Understanding Personal Data Externalization and Self-Tracking for Neurodivergent Individuals}

\author{Tanya Rudberg Selin}
\affiliation{%
  \institution{IT University of Copenhagen}
  \city{Copenhagen}
  \country{Denmark}}
\email{tanya.rudbergselin@gmail.com}

\author{Danielle Unéus}
\affiliation{%
  \institution{Uppsala University}
  \city{Visby}
  \country{Sweden}}
\email{danielle@uneus.se}

\author{Søren Knudsen}
\affiliation{%
  \institution{IT University of Copenhagen}
  \city{Copenhagen}
  \country{Denmark}}
\email{mail@sorenknudsen.com}

\renewcommand{\shortauthors}{Rudberg Selin et al.}

\begin{abstract}
We examine how neurodivergent individuals experience creating, interacting with, and reflecting on personal data about masking. Although self-tracking is often framed as enabling self-insight, this is rarely our experience as neurodivergent individuals and researchers. To better understand this disconnect, we conducted a two-phase qualitative study. First, a workshop where six participants with autism and/or ADHD crafted visual representations of masking experiences. Then, three participants continued by designing and using personalized self-tracking focused on unmasking over two weeks. Using reflexive thematic analysis of activities and interviews, we find that self-tracking imposes substantial interpretive and emotional demands, shaped by context-dependencies that challenge assumptions in self-tracking. We also find that facilitated sharing of experiences might validate emotional responses and support reflection. We identify three emotional dimensions that shape engagement with personal data in a working model of emotion in self-tracking, and discuss implications for designing self-tracking and reflective practices that incorporate peer support and better account for context and emotional labor.
\end{abstract}



\maketitle

\section{Introduction}

Self-tracking tools and personal informatics systems are widely used to ``collect personally relevant information,'' leading to ``self-knowledge''~\cite{li2010stagebased}. Such tools and systems are frequently described as supporting richer self-awareness, reflection, and insight into personal experience, with potential
benefits for wellbeing~\cite{barker-canler2024flexible}. From the basis of self-reflection, these technologies have been explored across different application domains, such as physical activity, food intake, productivity, social interactions, and mental health~\cite{epstein2020mapping}. 

Within this body of work, research has largely
focused on domains (i.e., what phenomena are tracked) and how tracking is implemented through particular systems and representations (i.e., how phenomena are tracked). 
Far less attention is paid to 
the people expected to engage with tracking practices within these domains. 
As a result, the benefits promised by self-tracking may be undermined when individual skills, capacities, and lived contexts are left unexamined. 
For example,~\citet{epstein2020mapping} examine ``what data domains [...] the personal informatics literature focus on,'' but largely sidestep how individual competencies shape
engagement with self-tracking. 
In this sense, the knowledge, skills, and practices of those meant to
benefit from self-tracking risk being treated as implicit or taken for granted.
Notably, this gap persists even in research that explicitly aims to support individuals with
situationally atypical cognitive function, as reported by \citet{homewood2024unanticipated,homewood2025cripping}.
This suggests a
broader tendency within personal informatics and HCI research to under-theorize how cognitive diversity intersects with data practices.

In this paper, we examine self-tracking from the perspective of neurodivergent lived experience, focusing on autistic and ADHD individuals. Prior work has emphasized that design
assumptions in HCI often reflect dominant cognitive norms, motivating calls for design approaches
that are sensitive to diverse cognitive patterns~\cite{fabri2019special, spiel2022adhd, azuka_inclusive_2024, wu_data_2023}. 
These recommendations highlight the need for flexible and personalized self-tracking systems that can account for individual complexity, particularly in personal data practices aimed at understanding and managing mental health~\cite{kim_toward_2019,simm_anxiety_2016,murnane2016selfmonitoring,spiel2022adhd}. 
At the same time, research has identified accessibility barriers that autistic adults face when using standard survey instruments~\cite{nicolaidis2020creating,uglik-marucha2026fit}, and shown that conventional approaches to visualizing emotion often fail to capture the embodied and context-dependent nature of autistic emotional experience~\cite{zolyomi_social-emotional-sensory_2021}. 
All of these examples highlight the value of understanding neurodivergent perspectives on personal data practices. 
Yet, despite growing attention to neurodiversity in HCI, research that centers neurodivergent lived experience and meaningfully involves neurodivergent people in design processes remains sparse~\cite{spiel2022adhd}.

As neurodivergent authors and researchers with ADHD and autism, we approach this work from within these experiences.
While self-tracking is often presented as something people can `just do,' this framing does not align with our own experiences. 
Although we possess the technical skills typically assumed to support self-tracking (e.g., data analysis), we continue to encounter challenges in attempting such practices. 
While prior work has documented challenges in self-tracking~\cite[e.g.,][]{choe2014understanding,epstein2015lived}, these challenges differ in important ways from our experiences. 
We notice that personal informatics systems often presume capacities that many neurodivergent individuals struggle to sustain, including formalizing ambiguous experiences, maintaining stable routines, and fitting internal states into discrete categories. 
These presumptions reflect cognitive norms that do not align with neurodivergent modes of experiencing and interpreting the world.

To examine these tensions, we focus on neurodivergent masking and unmasking in autistic and/or ADHD individuals.
Masking refers to modifying or suppressing behaviors to `fit in' within neuronormative contexts~\cite{pearson2021conceptual}.
The concept is most extensively theorized in autism research, where masking has been linked to distress, identity disruption, and poor mental health~\cite{russo2018costs,evans_what_2024}, and it is increasingly discussed in relation to ADHD~\cite{vanderputten2024camouflaging,ai2024dimensional}.
Because masking is ambiguous, context-dependent, and emotionally complex~\cite{ai2022reconsidering}, it poses particular challenges for conventional self-tracking practices and can amplify the challenges we investigate. 
Unmasking (the process of recognizing and understanding one’s masking habits, and of reclaiming more authentic and self-directed ways of being) is often described as central to neurodivergent wellbeing, yet individuals often lack supportive tools or environments to facilitate this process~\cite{harmens2022quest, raymaker2020having, mcfarlane_qualitative_2025}.
And, since unmasking relies heavily on self-reflection and self-knowledge~\cite{pearson2021conceptual}, personal data practices could offer valuable support. 
These reflections inform our focus and guiding research question: \textit{How do neurodivergent individuals experience creating and reflecting on personal data that represents ambiguous, complex behaviors such as masking?}

In response, we conducted a two-phase study with six neurodivergent participants to explore masking through personal data engagement, group reflections, and individual interviews. Our study phases respectively focused on \textit{externalization} (creating visual representations for immediate reflection) and \textit{self-tracking} (sustained logging over time). In Phase I, we hosted a workshop in which six participants externalized their lived experiences with masking, using a questionnaire as well as drawing and writing prompts. Participants then reflected in group discussions. In Phase II, three participants designed personalized self-tracking approaches and tracked masking for 1-2 weeks.  
Both phases were followed by interviews about their experiences with these approaches. We present our findings in two parts: 
1) participants' externalizations and self-tracking approaches, and 2) participants' reflections on externalization and self-tracking, which we present across four themes.
Our work demonstrates how complex neurodivergent contexts contribute to challenges in self-tracking. We identify emotional dimensions in personal data engagement as well as opportunities for responding to these aspects through shared and peer-supported reflection. 
Based on our work, we make three main contributions: 

\textbf{(1) We show that self-tracking might benefit from considering social and emotional aspects.} Through this, we reframe self-reflection and self-insight as activities and learning that are supported and informed by peers.
Our findings demonstrate emotional challenges related to personal data practices: generalization and quantification provoked overthinking, while commitment to longer-term tracking (unrealistically) demanded stable and available mental resources. 
We discuss participants' experiences with self-tracking methods and how they contributed senses of: 
\textit{emotional weight} (e.g., immediate stress from overthinking; Section \ref{sec:emotional-weight}), 
\textit{emotional self-reflection} (e.g., discomfort from insights surfaced; Section \ref{sec:emotional-self-reflection}), and
\textit{emotional burden} (e.g., pressure of maintaining routines; Section \ref{sec:emotional-burden}).

\textbf{(2) We describe how shared and peer-supported reflection might transform the emotional challenges} and positively affect self-tracking processes. Experiences that triggered anxiety when faced alone became validating when shared and recognized collectively. Based on this, we discuss the potential for a stronger focus on peer support in the design of self-tracking approaches, devices, and systems (Section \ref{sec:peer-support}).
Such designs might ease the emotional weight and burden of self-tracking, as well as scaffold reflection and self-understanding.

\textbf{(3) We exemplify how accessible study participation can be enabled by careful and richly informed study design,} rooted in our own experiences as neurodivergent individuals and researchers (Section \ref{sec:study-design}). These methodological approaches might inform other neurodiversity-focused HCI research and, potentially, HCI research with other marginalized groups.

\section{Background and Related Work}

\subsection{Neurodivergent Experience}
Neurodiversity refers to the natural variation in human cognition, affect, and perception. It challenges deficit-based models that treat autism, ADHD, and other conditions as pathologies~\cite{goldberg2023unraveling}. The term emerged in the 1990s through autistic self-advocacy and disability activism~\cite{goldberg2023unraveling,shah2022neurodevelopmental}. It has since informed both political movements and the academic field of neurodiversity studies, which interrogates neuronormative assumptions and centers lived experience as knowledge~\cite{rosqvist2020neurodiversity}.

Engaging in adaptive behaviors, and thus masking, has been described as a form of impression management that most people engage in to some extent~\cite{ai2022reconsidering,pryke-hobbes2023workplace, livingston2019compensatory}. However, for neurodivergent individuals, masking is rarely discretionary. It is instead a compulsory response to neuronormative pressures, and this enforced dimension renders masking particularly damaging to wellbeing~\cite{pearson2021conceptual}. Masking often results from internalized stigma or experiences of rejection, and consequently becomes automated and performed by habit~\cite{cage2019understanding}. Research suggests that neurodivergent masking often damages mental, physical, and emotional wellbeing for both adolescents and adults~\cite{russo2018costs, halsall2021camouflaging, raymaker2020having,mantzalas2022what}. However, when masking is consciously and situationally chosen, it may function as an adaptive practice that supports safety, access, and a sense of agency~\cite{kritika2025ultimately}. This highlights the importance of unmasking as a process of developing awareness of, and agency over, one’s masking practices given its implications for neurodivergent wellbeing~\cite{pearson2021conceptual}.

Neurodiversity studies emphasize that neurodivergent identities are highly diverse, shaped by intersecting cultural, social, and personal factors rather than reducible to a single label~\cite{hillary2020neurodiversity}. This diversity entails intertwined vulnerabilities and strengths that shape how individuals navigate neuronormative environments. For instance, research highlights that autistic and ADHD populations demonstrate creativity and responsiveness to positive environments, and experience rich and nuanced emotions~\cite{sedgwick2019positive,schippers2022qualitative, zolyomi_social-emotional-sensory_2021}. At the same time, research suggests common challenges including fluctuating emotions, energy depletion, difficulties with task completion despite genuine interest~\cite{soler-gutierrez2023evidence} and challenges relating to ambiguity~\cite{furnham1995tolerance, boulter2014intolerance}. Neurodivergent individuals also face heightened criticism, both externally through neuronormative contexts and internally through self-judgment, including overthinking, perfectionism, and catastrophizing~\cite{beaton2020selfcompassion,macalister2024autism}. Together, these dynamics can increase reliance on masking as a means of managing social evaluation~\cite{kritika2025ultimately}, while also heightening risks for depression and anxiety~\cite{beaton2020selfcompassion, oskouei2025mediating}. Practices that foster self-acceptance, such as positive self-talk and unmasking, can help buffer negative effects~\cite{parker2011examination, pearson2021conceptual}.

HCI research is increasingly engaging with neurodiversity~\cite[e.g.,][]{spiel2022adhd,kritika2025ultimately}.
For instance, Crip HCI advocates for a more in-depth integration of critical disability studies within HCI and emphasizes the need for greater involvement of disabled individuals, including neurodivergent people, in design processes~\cite{williams2021articulations}. 
However, HCI research that is grounded in neurodivergent lived experience and led by neurodivergent perspectives, rather than interpreting neurodiversity from the outside, remains relatively sparse~\cite{denhouting2019neurodiversity,spiel2022adhd,kritika2025ultimately,gollasch2023designing}.

\subsection{Self-Tracking and Personal Data Engagement}
Self-tracking is commonly used to support self-reflection, gain insights, and potentially, to change behaviors~\cite{epstein2020mapping}. 
For example, self-tracking is used in research interventions, as well as in clinical settings, including behavior therapy~\cite{choe2014understanding}. It is also increasingly common for individuals to collect and reflect on their personal data for self-understanding and self-improvement~\cite{huang2015personal}. 
While self-tracking technologies gain increasing popularity, dominant frameworks and existing tools often overlook the embodied, contextually-situated nature of neurodivergent emotional experience~\cite{nicolaidis2020creating, spiel2022adhd,zolyomi_social-emotional-sensory_2021}. Recent work by Alrøe and Krogh~\cite{alroe2025decentering} on easing ``the burden of conducting autistic autoethnography'' is a notable exception in this landscape.

The stage-based model proposed by Li et al.~\cite{li2010stagebased} has framed research in personal informatics systems and made a strong impact on later self-tracking work. Relevant to our study, their questionnaire respondents reported challenges related to remembering to collect data and to accuracy during the data collection phase.
Though some later work discussed `forgetting' in relation to device maintenance (e.g., forgetting to wear or charge tracking devices~\cite{epstein2016reconsidering}) and how people often stop using tracking over time (e.g., ``the lapsing stage''~\cite{epstein2016reconsidering}), our work relates to forgetting to log and challenges that impede tracking from the outset. 

In relation to food tracking, Karkar et al.~\cite{karkar2017tummytrials} describe participants' frustration with strict cutoff times for reporting symptoms, as well as lack of flexibility in which symptoms could be tracked and their severity, but offer limited guidance in response. At the same time, examining the practices of ``quantified selfers,'' Choe et al.~\cite{choe2014understanding} highlight a common pitfall of ``tracking too many things.'' Yet, they note that early in a tracking practice, high motivation can make this manageable and even useful for deciding which items to keep, but over time it often leads to ``tracking fatigue.'' Acknowledging the need for understanding individual differences, Paay et al.~\cite{paay2015understanding} discuss the use of contextual reminders in relation to apps for smoking cessation.
More broadly, such concerns are partly acknowledged in discussions of `lived informatics'~\cite{epstein2015lived}.

Thudt et al.~\cite{thudt2018selfreflection} explore the use of physicalizations for self-reflection. In their work, participants constructed data physicalizations to reflect on personal topics over several weeks. Their results show how personalization and integration into daily routines supported deeper reflection. 
They also found that data collection, visualization, and reflection were interwoven: the practice of constructing physicalizations prompted reflection for the participants. Thus, their work challenges the common description of reflection as a subsequent step to data collection~\cite[e.g.,][]{li2010stagebased}, showing potential for reflection throughout the full process of self-tracking. In our study, we explore this potential through \textit{externalization} of personal data: externalizing one's internal experiences, including behaviors, emotions, or thoughts, through external representations~\cite{larkin1987why, walny2011visual}.

\subsection{Neurodiversity and Data}
Data visualization research has suggested inclusive and personalized design approaches to ensure neurodivergent individuals' active contributions in shaping solutions to meet their preferences and needs~\cite{wu2021understanding,wu2024our}. For example, Wu et al.~\cite{wu2024our} conducted co-design workshops with individuals with intellectual and developmental disabilities, combining self-discovery with data representation. They used everyday materials and contexts to make data more accessible and personally relevant to participants.

Zolyomi and Snyder's work on social-emotional-sensory design for affective computing provides crucial insights into how neurodivergent adults engage with representations of emotion~\cite{zolyomi_social-emotional-sensory_2021}. Their participants described emotions as embodied, co-constructed within social contexts, and entangled with sensory experiences. The study proposes a ``social-emotional-sensory design map'' as an alternative framework for affective computing. Rather than attempting to classify emotions into predetermined categories, their framework acknowledges the constructed, personalized nature of emotional experiences.

Neurodivergent masking exemplifies these highly personal emotional experiences. Masking may involve both physical and psychological adaptations, with research highlighting examples including hiding personal quirks and emotions, altering body language, monitoring facial expressions, and avoiding discussion related to oneself~\cite{cook2022selfreported,hull2019development, kritika2025ultimately}. As such, masking is highly context-dependent and complex. Using impression management (IM) to describe masking and highlight these complex phenomena, Ai et al.~\cite{ai2022reconsidering} ask: ``What are the optimal ways to measure IM that consider its dynamic, iterative, context-dependent, and transactional nature, and capture the unique aspects experienced by autistic and other neurodivergent people?'' This question calls for future research on this topic. 

Currently, a widely used tool to measure masking is the Camouflaging Autistic Traits Questionnaire (CAT-Q)~\cite{hull2019development}, designed for adults to self-assess their masking behaviors. It consists of 25 claims about masking (e.g., ``In social situations, I feel like I am pretending to be `normal'''). Responses are given on a 7-item Likert scale. The CAT-Q is a recognized contribution to the neurodiversity community and, with more than 500 citations, is well-cited in academia. However, the use of Likert scales with imprecise response options might present as an accessibility barrier for autistic adults~\cite{nicolaidis2020creating}. Due to the complexities inherent in neurodivergent masking, we assume that a questionnaire cannot fully capture these experiences, nor is it intended to do so. Clearly, related work demonstrates the gap between established self-tracking practices and what we ourselves experience as neurodivergent individuals. 

\section{Approach}
To examine how neurodivergent individuals experience collecting and reflecting on personal data about masking, we engaged neurodivergent participants in two study phases.
Throughout, we adopted an exploratory approach and therefore engaged participants in open-ended and varied activities for personal data engagement and reflection. We chose this approach for two reasons: to broadly examine the possibilities in an understudied area (combining self-tracking and neurodivergent masking) and to allow our participants to shape their own path in our study.

\subsection{Study Design}
\label{sec:study-design}
\begin{figure*}[tb!]
    \centering
    \includegraphics[width=\linewidth]{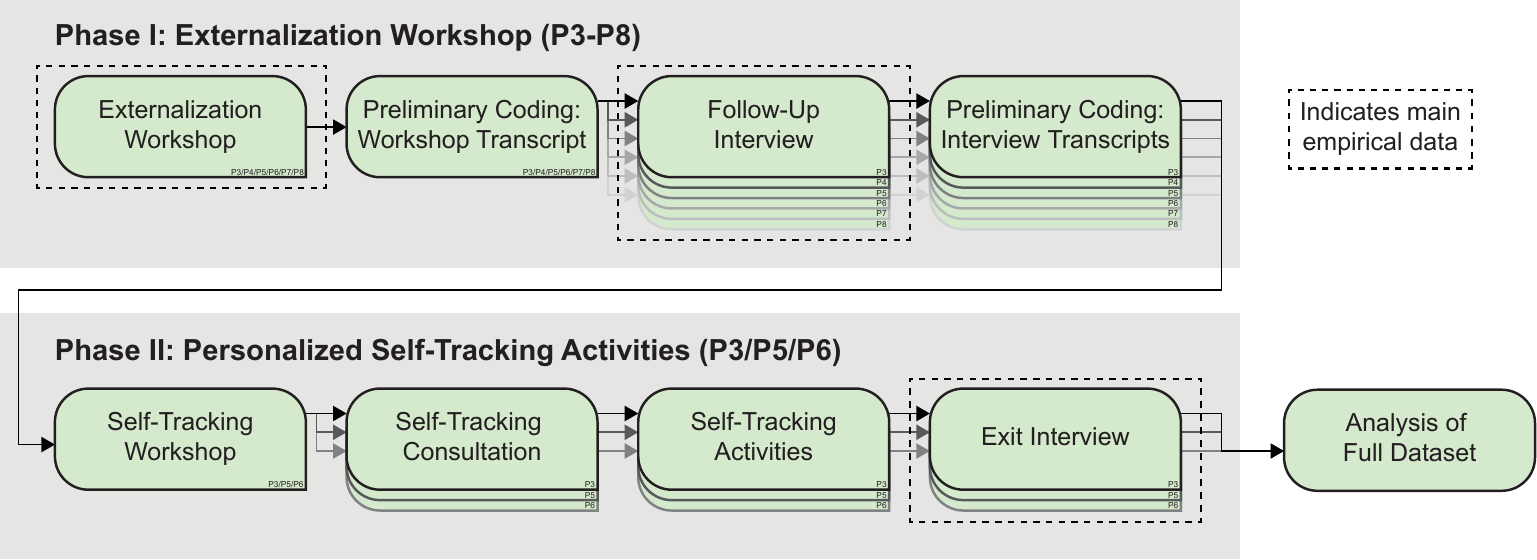}
    \caption[]{Timeline of the study, consisting of two main phases (each comprised of four stages) and a final analysis stage. 
    Each stage is represented as a rounded rectangle, a title and participants present for the activity.
    Arrows connect the nine stages, indicating their order, while the three stages that comprise the main empirical data are framed by dashed rectangles.
    }
    \Description{
A timeline figure showing two study phases (Phase I and Phase II), each with four stages represented as green rounded rectangles. Phase I (Externalization Workshop) involved six participants (P3, P4, P5, P6, P7 and P8) in four stages: Externalization Workshop, Preliminary Coding of Workshop Transcript, Follow-Up Interview, and Preliminary Coding of Interview Transcripts. Phase II (Personalized Self-Tracking Activities) involved three participants (P3, P5, P6) in four stages: Self-Tracking Workshop, Self-Tracking Consultation, Self-Tracking Activities, and Exit Interview. Overlapping rectangles indicate stages in which participants took part individually (Follow-Up Interview, Self-Tracking Consultation, Self-Tracking Activities, and Exit Interview). A final Analysis of Full Dataset stage follows Phase II. Dashed rectangles indicate the three main empirical data collection stages: Externalization Workshop, Follow-Up Interview, and Exit Interview.
}
    \label{fig:process-timeline-1}
\end{figure*}

We designed our study with two complementary phases, engaging participants in \textit{externalization} (Phase I) and \textit{self-tracking} (Phase II). We varied the social context of reflection between shared and individual settings.
Figure \ref{fig:process-timeline-1} presents an overview of the study phases.

Phase I explored in-the-moment data creation through externalization. We asked participants to complete the CAT-Q~\cite{hull2019development} and engaged them with drawing and writing prompts to create externalizations representing masking. We refer to these activities as \textit{externalization exercises} and participants' outputs as \textit{externalizations} or \textit{representations} (of masking). Although not commonly associated with such descriptions, we view the act of completing a questionnaire as externalization. 
This interpretation and methodological choice is inspired by Larkin and Simon~\cite{larkin1987why}: 
``When they are solving problems, human beings use both internal representations, stored in their brains, and external representations, recorded on a paper, on a blackboard, or on some other medium.'' We also draw on visual thinking and data sketching~\cite{walny2011visual, walny2015exploratory}: the act of forming a mental image of an idea (internal visualization) and sketching to externalize it (external visual representation of the idea), supporting spontaneous reflection and communication.

The first phase centered on group discussions in a workshop. This design choice drew inspiration from our own experiences with group approaches in neurodivergent communities and clinical mental health contexts. We note agreement in literature relating to group therapy~\cite[e.g.,][]{geng2025when}. Moreover, we notice related benefits of shared reflections within households in Thudt et al.'s physicalization study~\cite{thudt2018selfreflection}. Our participants completed the CAT-Q~\cite{hull2019development} before the workshop and discussed these experiences during it. We chose to include the CAT-Q for two reasons. First, it is a widespread tool for self-assessment of masking. Second, even though emotional experiences are difficult to quantify (i.e., for both neurotypical and neurodivergent people)~\cite{lottridge2011affective}, the need to do so is common in self-tracking. Unlike typical research that uses such instruments to assess or categorize participants, we never asked participants to share their scores. Instead, we positioned the CAT-Q --- along with the drawing and writing prompts --- as externalization activities that could serve as a foundation for reflection. Participants reflected both on what these externalizations represented about their masking experiences and on the process of creating them.

Phase II complemented these immediate reflections with sustained self-tracking over one to two weeks. Self-tracking literature emphasizes consistent, longer-term tracking for producing insights and supporting reflection. However, while such literature often focuses on behavioral change~\cite[e.g.,][]{lee2017selfexperimentation}, we were inspired by the focus on reflection articulated by Thudt et al.~\cite{thudt2018selfreflection}. Focusing on reflection rather than change, we saw less need for the extended consistency typically emphasized in self-tracking. We wanted to explore the potential reflective benefits of sustained engagement with personal data about masking while ensuring the study remained accessible and sustainable for neurodivergent participants. Thus, 
we deliberately designed the self-tracking period to be short and flexible to reduce barriers to participation.

\subsubsection{Contextual Attentiveness}
We designed our study with attention to the context we explored. For instance, we recognize that neurodivergent identities and preferences are highly diverse~\cite{hillary2020neurodiversity, denhouting2019neurodiversity}. We therefore sought to highlight the individuality and personal experiences of our participants. To achieve this, we conducted individual interviews and consultations with them throughout the study. While we provided structure for the study approach, we allowed participants to shape their own path. For instance, we set self-tracking as a core activity but invited participants to choose the specific aspects to track and their approaches (as in Thudt et al.~\cite{thudt2018selfreflection}). We also sought to accommodate diverse preferences. Participants could choose to create externalizations and complete self-tracking either digitally or by hand, and could select their preferred interview method. We collected and considered their feedback continuously throughout the study and before continuing to a new study phase.

It is common for ADHD and/or autistic individuals to experience sensory sensitivities as well as fluctuating emotions and energy~\cite{soler-gutierrez2023evidence, zolyomi_social-emotional-sensory_2021}, and masking has been specifically linked to exhaustion~\cite{pryke-hobbes2023workplace}. We found this important to consider throughout the research process. To avoid overwhelming our participants, we started with a time-restricted workshop and invited participants to join self-tracking activities in a later phase. We thus prioritized the early stages to let participants engage with masking as an ambiguous phenomenon and explore personal data practices while lowering the demand for commitment and cognitive load. To further reduce participation barriers, we conducted the study online. Workshops were facilitated on video call (Zoom) and individual interviews were carried out via video call, phone call, or e-mail. As Halsall et al.~\cite{halsall2021camouflaging} note in relation to masking, neurodivergent individuals often find they can be themselves and express themselves more freely in familiar contexts such as their home. Enabling participants to join from their home environments thus sought to lower the mental demands of participation, reduce logistical barriers, and accommodate sensory sensitivities and fluctuating energy levels.


\begin{table*}
  \caption[]{Participants are listed with their anonymized ID, which we use to refer to them throughout. All participants share Swedish as their native language.}
  \label{tab:participants}
  \begin{tabular}{l l l l l l l}
\toprule
    \textbf{Participant} &
    \textbf{Gender} &
    \textbf{Age} &
    \textbf{Occupation} &
    \textbf{Education} &
    \multicolumn{2}{l}{\textbf{Phases}} \\

    &
    &
    &
    &
    & \textbf{I} &
    \textbf{II} \\

\midrule

    \textbf{P3} &
    Male &
    31 &
    Unemployed &
    MSc, Field Biologist &
    \huge\raisebox{-0.75pt}{\textbullet} &
    \huge\raisebox{-0.75pt}{\textbullet} \\
    
    \textbf{P4} &
    Female &
    31 &
    Electrician &
    VocEd, Electrician & 
    \huge\raisebox{-0.75pt}{\textbullet} &
    \\

    \textbf{P5} & 
    Female &
    30 &
    Electrician (apprentice) &
    VocEd, Electrician & 
    \huge\raisebox{-0.75pt}{\textbullet} &
    \huge\raisebox{-0.75pt}{\textbullet} \\
    
    \textbf{P6} &
    Non-binary & 
    30 & 
    Catering Chef &
    Secondary Education & 
    \huge\raisebox{-0.75pt}{\textbullet} &
    \huge\raisebox{-0.75pt}{\textbullet} \\
    
    \textbf{P7} & 
    Male & 
    30 &
    On sick leave &
    Not reported & 
    \huge\raisebox{-0.75pt}{\textbullet} &     
    \\
    
    \textbf{P8} &
    Male & 
    28 &
    Self-Employed, TV \& Media & 
    VocEd, Locomotive Engineer & 
    \huge\raisebox{-0.75pt}{\textbullet} &     
    \\

\bottomrule
  \end{tabular}
\end{table*}

\subsubsection{Participants}
We recruited participants through our networks. 
Six participants (3 male, 2 female, 1 non-binary, aged 28 to 31 years) joined Phase I, all identifying as neurodivergent and with diagnoses of ADHD, autism, or both.  
All six participants were invited to continue into Phase II.
Four initially accepted, but one withdrew before Phase II began.
Hence, three participants (1 male, 1 female, 1 non-binary) completed Phase II.
See Table \ref{tab:participants}.

\subsubsection{Positionality}
As researchers, we approach this work from neurodivergent perspectives.  We study the communities of which we are members ourselves, with lived experiences of ADHD and autism. Our positionality shaped how we designed the study, facilitated workshops, and analyzed data. 
In designing our study, we drew on experience as participants and organizers of online and in-person events by and for neurodivergent people, and about neurodiversity.
During workshops, we drew on our own experiences with masking to build rapport and shared understanding with participants. 
In our analysis, our lived experiences influenced what we attended to in participants' accounts, how we interpreted their meanings, and how we chose to represent findings. 
We see our positionality as essential to constructing analysis from an insider's perspective~\cite{denhouting2019neurodiversity} while recognizing that our interpretations reflect our particular positions within these communities rather than universal neurodivergent experience. 
We return to this aspect in our discussion (Section~\ref{sec:strengths-limitations}).

\subsection{Phase I: Externalization Workshop}
We hosted a three-hour workshop with all six participants and subsequently collected their feedback in follow-up interviews. The workshop compromised three one-hour stages:

\textbf{(1) Quantification and Questionnaires.} 
A group discussion centered on experiences with questionnaires and quantification. 
This discussion was guided by three open-ended questions, asking participants about their 
a) general experiences with quantitative reporting of mental health, 
b) specific experiences using a Likert scale to externalize masking in the CAT-Q~\cite{hull2019development}, and 
c) overall experiences with completing the CAT-Q.

\textbf{(2) Drawing and Writing Prompts.} 
Two iterations of externalization exercises. 
In each iteration, participants spent 10 minutes individually creating representations of their masking experiences --- first through drawing, then through writing --- before gathering in two smaller groups (three participants and one facilitator in each) for 20 minutes to share and discuss their externalizations. 
To level with participants, we participated in these exercises alongside them and shared our own externalizations in group discussions.

\textbf{(3) Reflection and Ideation.} 
We convened in plenum to reflect on the workshop experience and ideate on broader possibilities for using personal data to support unmasking. 
This discussion was guided by four open-ended prompts concerning: 
a) what worked well and less well in the workshop, 
b) ideas for similar future workshops, 
c) reflections on unmasking, and 
d) ways in which unmasking might be supported.

After the workshop, we conducted individual follow-up interviews with participants, who were free to respond by email (P3, P5, P7) or phone call (P4, P6, P8). 
Our follow-up interview guide was informed by workshop findings and the Critical Incident Questionnaire (CIQ)~\cite{keefer2009critical}, which we have previously used successfully in our teaching and reflective practices. 
We asked participants for feedback on the workshop and further reflections on unmasking. 
These interviews also served to invite participants to Phase II, which three of them (P3, P5, and P6) chose to complete.

\subsection{Phase II: Personalized Self-Tracking Activities}

We engaged three participants (P3, P5, and P6) in personalized self-tracking activities and conducted exit interviews in which they reflected on these experiences.

\textbf{(1) Self-Tracking Workshop.} We organized a two-hour group workshop to prepare participants for self-tracking activities. We familiarized them with various data logging techniques and facilitated an ideation session to discuss potential approaches to self-tracking. Following, participants were given three days to reflect and begin developing their own personalized approaches.

\textbf{(2) Individual Consultations.} We contacted participants to finalize their approaches. We agreed with P6 to track for the full two weeks, while with P3 and P5 we planned a check-in after the first week to allow them to decide whether to continue. All participants were informed they could modify their self-tracking at any point if they found it difficult or inefficient and were asked to note any changes and the reasons for them.

\textbf{(3) Self-Tracking Activities.} 
Participants implemented their personalized self-tracking approaches, focusing on masking behaviors in contexts most salient to them (P3: social situations, P5: school, P6: work). 
After the initial week, P3 decided to stop tracking, while P5 wanted to continue but later revealed she had stopped after the first week. 
P6 tracked for two weeks as planned.

\textbf{(4) Exit Interviews.} 
We conducted semi-structured exit interviews with all three participants via phone (P3, P6) or video call (P5), lasting between 40 and 60 minutes. 
All interviews were conducted after participants finished tracking and explored their self-tracking experiences from multiple angles. 
We asked them to describe their tracking approaches, explain their initial goals and whether these were met, and reflect on any challenges or adaptations they made during the process. 
We asked participants about both practical aspects (ease of use, surprises, whether they shared their data with others) and reflections on what they learned about self-tracking as a practice, about masking, and about themselves. 
We also asked them to consider hypothetical alternatives (e.g., whether digital tracking could have been done physically) and what they would do differently in future tracking efforts.

\subsection{Data Collection and Analysis}
We analyzed recordings and transcripts from the externalization workshop and follow-up interviews (Phase I) as well as exit interviews (Phase II) using reflexive thematic analysis~\cite{braun2019reflecting}. We chose to focus on these data sources because they capture participants articulating their experiences in creating and reflecting on their personal data. We also collected participants' externalizations from Phase I and self-tracking logs from Phase II to contextualize their explanations during analysis, but we developed themes based on what participants said rather than what they created. All sessions (workshops, consultations, and interviews) were conducted in Swedish. The first author transcribed recordings and translated transcripts into English for analysis.

\subsubsection{Participant Flow and Data Collection}
All six participants (P3-P8) completed Phase I, participating in the externalization workshop and subsequent individual follow-up interviews. Three participants (P3, P5 and P6) completed Phase II, participating in a self-tracking workshop, individual consultations, self-tracking activities, and subsequent individual exit interviews.

During Phase I, the first author transcribed and coded recordings shortly after each session. This enabled a responsive study design. Following the externalization workshop, the first and second authors (who had facilitated the workshop) independently reviewed the workshop transcript, highlighting noteworthy moments before meeting to discuss implications for the follow-up interview guide. After follow-up interviews, the first author reviewed interview transcripts and identified common themes in participants' feedback. We specifically examined the transcripts of P3, P5 and P6 to accommodate their preferences in Phase II design.

During Phase II, we focused on assisting participants individually in creating and using personalized self-tracking rather than performing preliminary coding. The self-tracking workshop and individual consultations were not transcribed or analyzed, but observations from these activities helped inform our exit interview guide. The first author conducted individual exit interviews with participants after they had engaged in self-tracking (P3 after one week, P5 and P6 after two weeks).

\subsubsection{Thematic Analysis}
\begin{figure*}[tb!]
  \centering
  \fbox{\includegraphics[width=0.225\linewidth]{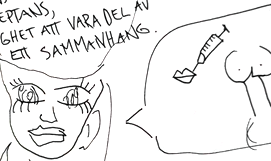}}
  \hfill
  \fbox{\includegraphics[width=0.225\linewidth]{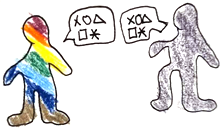}}
  \hfill
  \fbox{\includegraphics[width=0.225\linewidth]{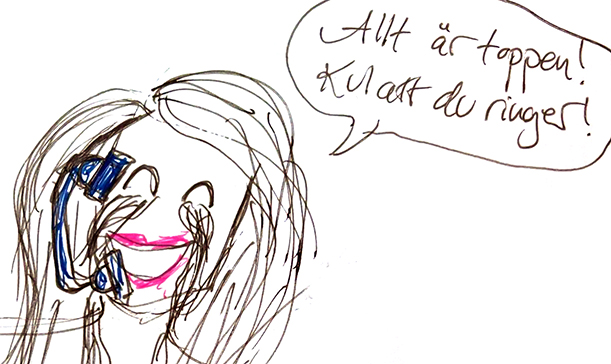}}
  \hfill
  \fbox{\includegraphics[width=0.225\linewidth]{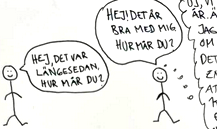}}
  \caption[]{
  Participants' externalizations (drawing prompt). All participants created self-portraits. The sketches shown here include those containing speech bubbles and are arranged along a visual-to-words spectrum (left to right). Relationships to other people appear across all drawings as a cross-cutting theme. From left to right: (1) P4 drew a self-portrait with a speech bubble containing symbols representing cosmetic surgery, symbolizing that ``as a woman you have to fit into a certain frame.'' (2) P4 depicted herself as a silhouette composed of seven colors, attempting to imitate a gray silhouette representing another person in the interaction; identical symbols in both speech bubbles emphasize this act of imitation. (3) P5 illustrated herself crying while answering the phone, with a speech bubble stating, ``Everything is great! It’s great that you are calling.'' (4) P4 drew a conversation between herself and another person who asks ``Hello, long time no see, how are you?'' and while P4 constructs a much longer answer in her mind, she responds, ``Hello, I’m good, how are you?''}
  \Description{The sketches rely on simple face drawings and stick figures to convey the points, which are mainly carried home through the speech bubbles as described in the figure caption.}
  \label{fig:visual-representations-1}
\end{figure*}

After completing all data collection, we conducted reflexive thematic analysis~\cite{braun2019reflecting} on the complete dataset (transcripts of externalization workshop, follow-up interviews, and exit interviews). 
We began by analyzing Phase II exit interviews, which explored individual experiences with self-tracking. 
To honor each participant's unique approach and experience, we analyzed the exit interviews individually (P3, then P5, then P6) before identifying themes across participants. The first author transcribed all exit interviews and translated transcripts to English, pairing Swedish and English versions side-by-side with visual examples of each participant's self-tracking setup and logs for context. The first and third authors then collaboratively coded each interview, building an understanding of each participant's experience. Subsequently, we identified and developed themes by grouping codes that reflected patterns of shared meaning across the three participants.

We applied this thematic framework to Phase I data, with the first author working through workshop and interview transcripts while remaining open to emergent themes that might refine or extend the framework.  
Through this iterative work, new themes emerged both within the individual study phases and across them.
This analysis culminated in drafting descriptions of themes (first author) and revising and restructuring these through three major revisions of our analysis (first and third author). Throughout, we engaged in reflexive discussion about our positionalities and interpretations, iteratively refining themes through dialogue about their meaning and how our own perspectives shaped our readings of the data.

We present our findings in the next two sections. In Section \ref{sec:findings-externalizations-and-approaches}, we present these primarily through participants' externalizations and self-tracking from the individual phases. Then, in Section \ref{sec:findings-participant-reflections}, we present the analytic themes synthesized across both phases.

\section{Findings: Externalizations and Self-Tracking Approaches}
\label{sec:findings-externalizations-and-approaches}

This section describes the externalizations and self-tracking approaches that participants created in our study. We first describe the externalizations that six participants created in Phase I, then the self-tracking approaches that three participants designed and used in Phase II. For each phase, we present participants' data practices and their experiences creating and reflecting on these representations. We then synthesize common themes across both phases in the following section.

\subsection{Phase I: Externalization}

We first demonstrate participants' externalizations (Phase I) and their experiences in crafting and reflecting on them.

\subsubsection{What Participants Created}
\begin{figure*}[tb!]
  \centering
  \fbox{\includegraphics[height=3.35cm]{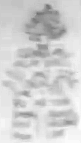}}
  \hfill
  \fbox{\includegraphics[height=3.35cm]{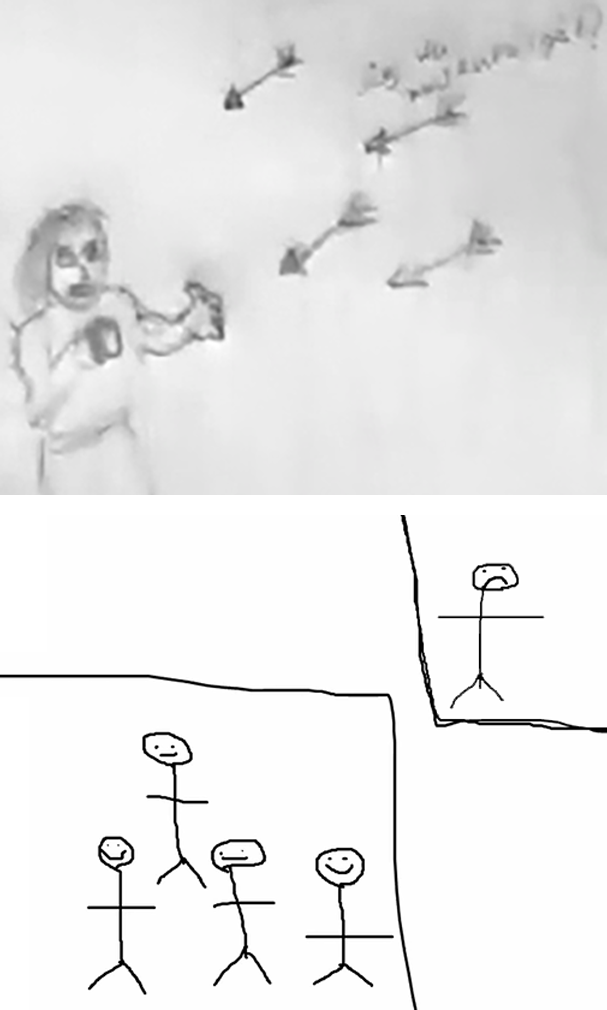}}
  \hfill
  \fbox{\includegraphics[height=3.35cm]{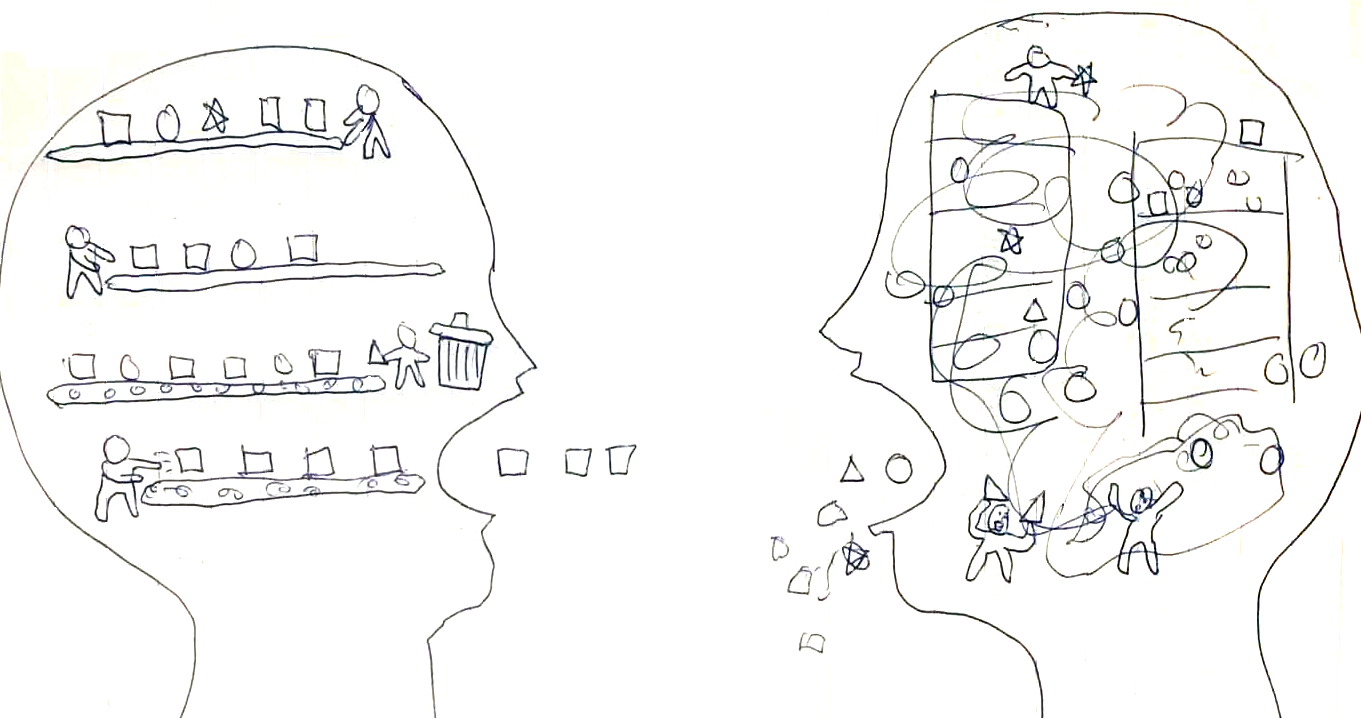}}
  \hfill
  \fbox{\includegraphics[height=3.35cm]{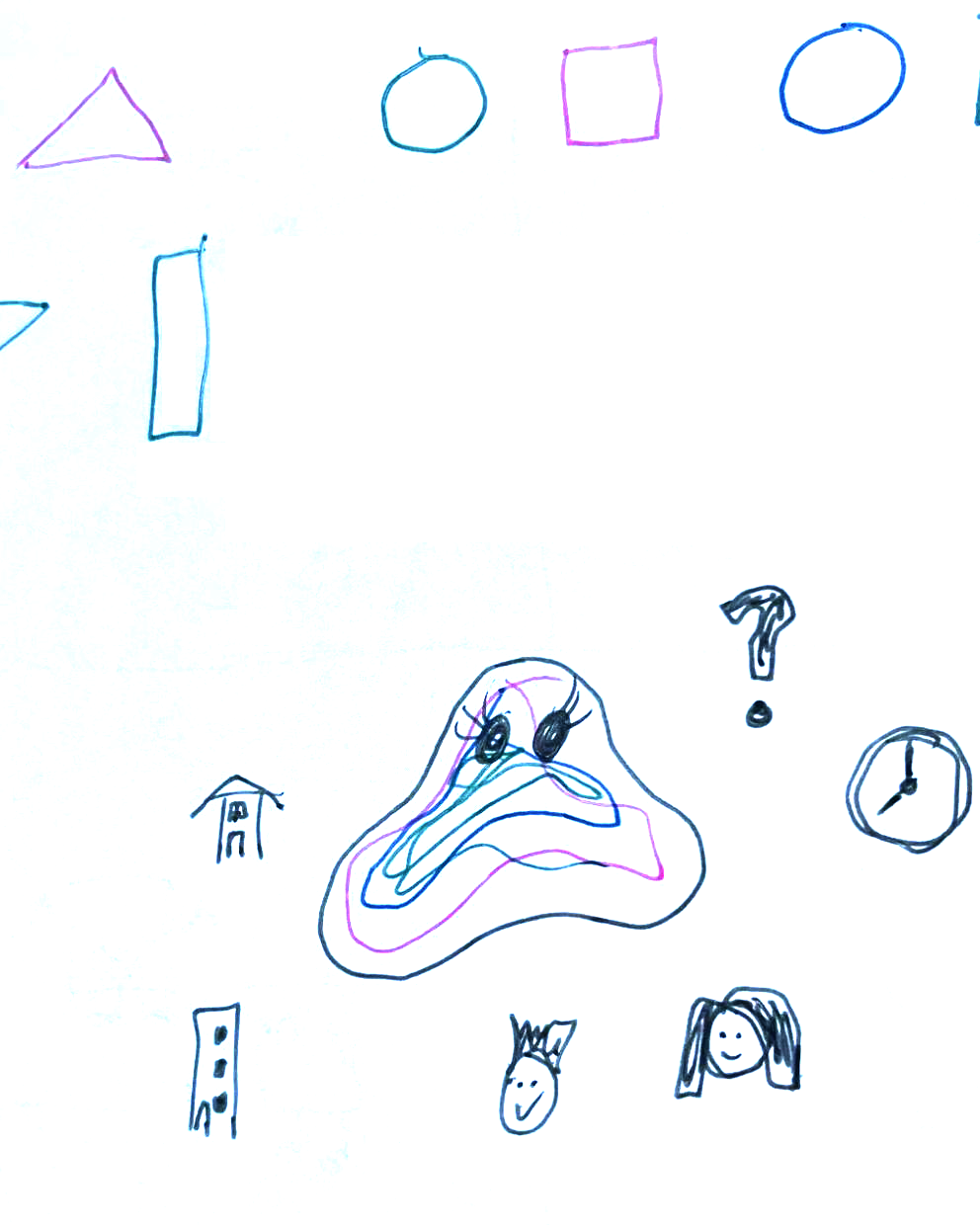}}
  \hfill
  \fbox{\includegraphics[height=3.35cm]{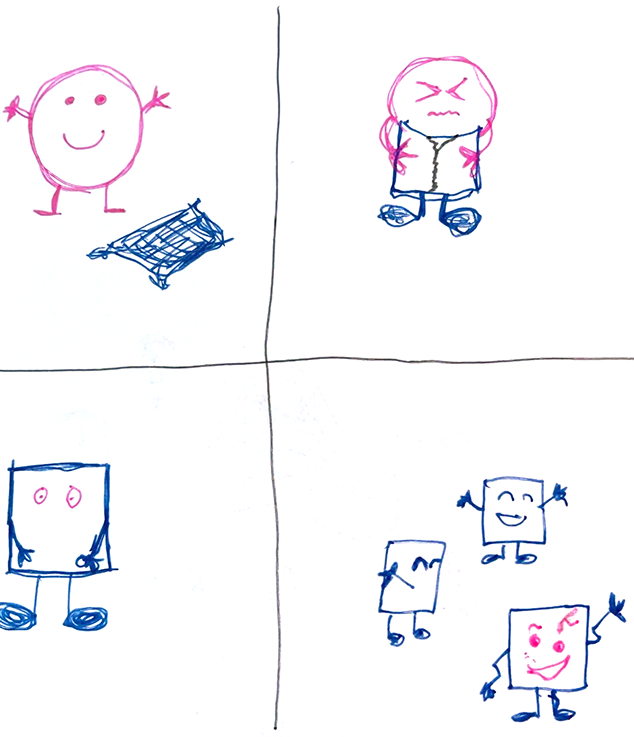}}
  \caption[]{Participants' externalizations (drawing prompt). These sketches present variations of self in relation to others. From left, top-to-bottom: (1) P6 sketched a silhouette of themselves formed by words and phrases that represent how they experience masking in various parts of their body. (2) P7 drew a self-portrait in which he stands still and makes a thumbs-up gesture while being attacked by arrows, which he explained as a metaphor for small talk ``attacking'' him. (3) P8 drew himself alone in a box with a sad expression, and four people smiling together in another box, describing these boxes as representing how he views masking as ``a barrier between myself and relating to other people.'' (4) P3 illustrated a conversation between himself and a neurotypical person, using a functional and a dysfunctional postal system as a metaphor to explain the processing within their respective and different minds. (5) P5 drew herself as a ``lump'' and various shapes as metaphors for masking behaviors, illustrating herself deciding ``which shape to squeeze into.'' (6) P5 drew a comic illustrating herself waking up as a round shape, squeezing into an uncomfortable corset to appear square, and arriving at school where everyone is square-shaped.}
  \Description{Besides the descriptions of sketches in the figure caption: sketches typically rely on stick figure drawings.}
  \label{fig:visual-representations-2}
\end{figure*}

\begin{figure*}[tb!]
  \centering
    \fbox{\includegraphics[height=2.7cm]{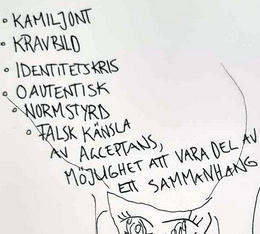}}
    \hfill
    \fbox{\includegraphics[height=2.7cm]{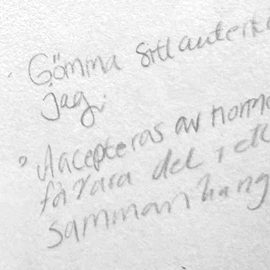}}
    \hfill
    \fbox{\includegraphics[height=2.7cm]{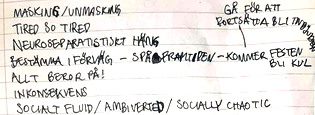}}
    \hfill
    \fbox{\includegraphics[height=2.7cm]{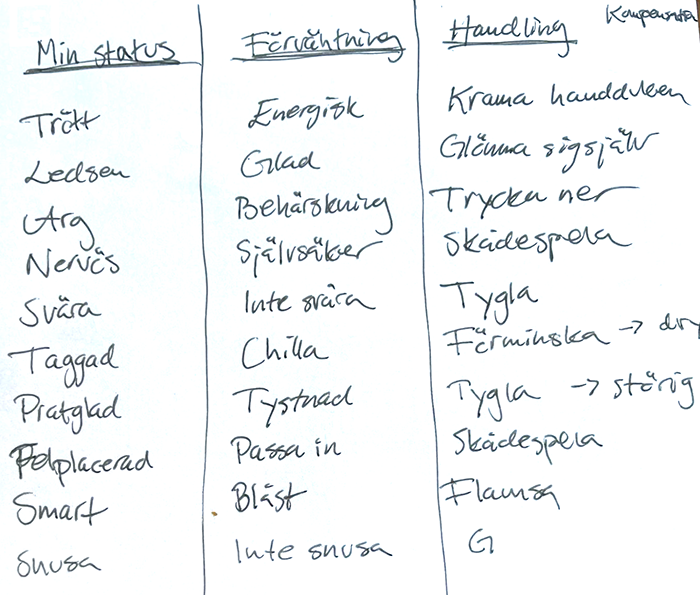}}
  \caption[]{Participants' externalizations (writing prompt). Participants used word lists, short phrases, annotated sketches, and tabular structures to describe masking. From left to right: (1) P4 extended her drawings from the earlier prompt with words placed in and around a thought bubble, expressing themes of identity, adaptation, and social belonging; (2) P4 added more verbal points next to her sketch; (3) P3 created a dense list of words and short phrases describing masking, exhaustion, and contextual variability; and (4) P5 organized words into a table contrasting current state, expectations, and actions.}
  \Description{The words in the sketches are in Swedish and thus mainly serve to illustrate the visual appearance as it is described in the figure caption.}
  \label{fig:visual-representations-3}
\end{figure*}

Figure \ref{fig:visual-representations-1}, \ref{fig:visual-representations-2}, and \ref{fig:visual-representations-3} show examples of participants' externalizations. We notice several themes across them. Although not prompted, all participants produced some form of self-portrait. These were often supplemented by visual metaphors for masking experiences such as: a dysfunctional postal system (P3), an uncomfortable corset (P5), and arrows representing pressure and expectation (P7). Many participants depicted masking as harmful to wellbeing: sad or strained facial expressions (P5, P8), images of physical discomfort or danger (P5, P7), and written phrases such as ``tired so tired'' (P3), ``inauthentic'' (P4), ``identity crisis'' (P4), and ``misplaced'' (P5). Another common thread was contextual adaptation, with visualizations situating masking in specific environments such as schools (P5), workplaces (P6, P7), and social events (P3).

\subsubsection{What Participants Said}

Participants described the process of creating visual representations of masking and discussing them in a group as valuable for both shared and individual reflections.

\begin{quote}
``It was rewarding to hear others share experiences that I could relate to. It's kind of my favorite way to talk about feelings or serious stuff. I feel validated without having to be the one to put things into words. And even though most of my friends have ADHD or autism, we don't often talk about masking or things from a broader perspective. So I felt good afterward, even though we didn't come up with any solutions to the problems.'' \textit{P3, follow-up interview (Phase I).}
\end{quote}

In follow-up interviews, all participants highlighted the shared discussions, particularly in the smaller breakout groups in externalization exercises, as a favorite aspect of Phase I. Participants expressed appreciation for both the self-recognition these discussions provided and the new perspectives they opened. P4 liked being reminded that ``there are more people experiencing and going through similar things.'' P5 noted that recognizing similarities with her own experiences was validating and affirming, while experiences dissimilar to hers contributed new reflective perspectives. P7 shared being surprised by ``how much I recognized myself in other participants’ experiences,'' and P6 valued gaining new perspectives on similar challenges. P3 expressed a surprise that ``something I thought sounded fluffy, like drawing and writing words, could lead to rewarding conversations.'' Similarly, P8 shared: ``I found it very interesting to hear other people's experiences on the subject [masking]. And that includes how, about other people’s drawings or, yes, how people had drawn things. It fascinated me a little.''

\subsection{Phase II: Self-Tracking}
Based on Phase I, three participants moved on to design and implement self-tracking.
Here, we present participants' self-tracking approaches and their experiences using them.

\begin{figure*}[t]
  \centering
  \includegraphics[width=\linewidth,page=1]{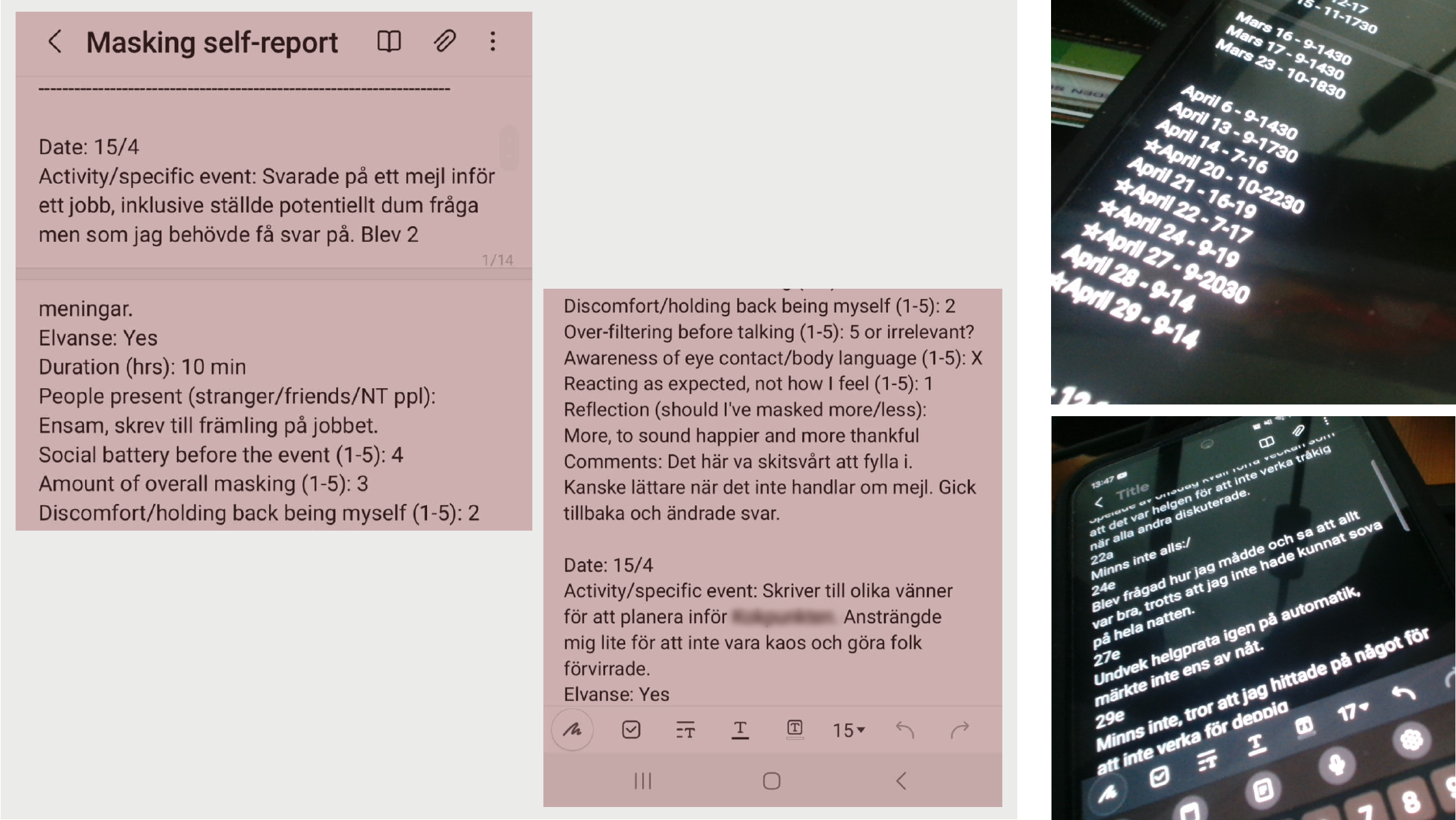}
  \caption[]{Excerpts of self-tracking materials from two participants. Left: A screenshot shown in two parts, displaying a self-tracking log from P3 reflecting on writing an e-mail to a coworker, including ratings of masking, discomfort, and reflective comments. Right: Two images from P6's phone showing self-tracking, combining marking of masking in their work schedule (right, top) and later written reflections in a notes app (right, bottom). Excerpts include reflections on automatic masking responses as well as comments on the difficulty of recollecting experiences for logging.}
  \Description{The figure shows two pairs of images, one pair created by P3 and one by P6. P3's pair is one screenshot divided in two to fit the page. In the screenshot, a title at the top reads `Masking self-report'. Below the title, two entries are shown from the same date, each entry with multiple lines of text that the participant would respond to, such as date, activity, Elvanse, duration, people present, social battery before event, and more. P6's pair shows two photos taken of a phone that shows daily entries. In the top photo, each entry is a line, starting with a star symbol (only for some), then date and time interval. We include full details in supplemental materials as they do not fit here.}
  \label{fig:self-tracking-designs-p3-and-p6}
\end{figure*}

\subsubsection{What Participants Created}
\begin{figure*}[t]
  \centering
  \includegraphics[width=\linewidth,page=2]{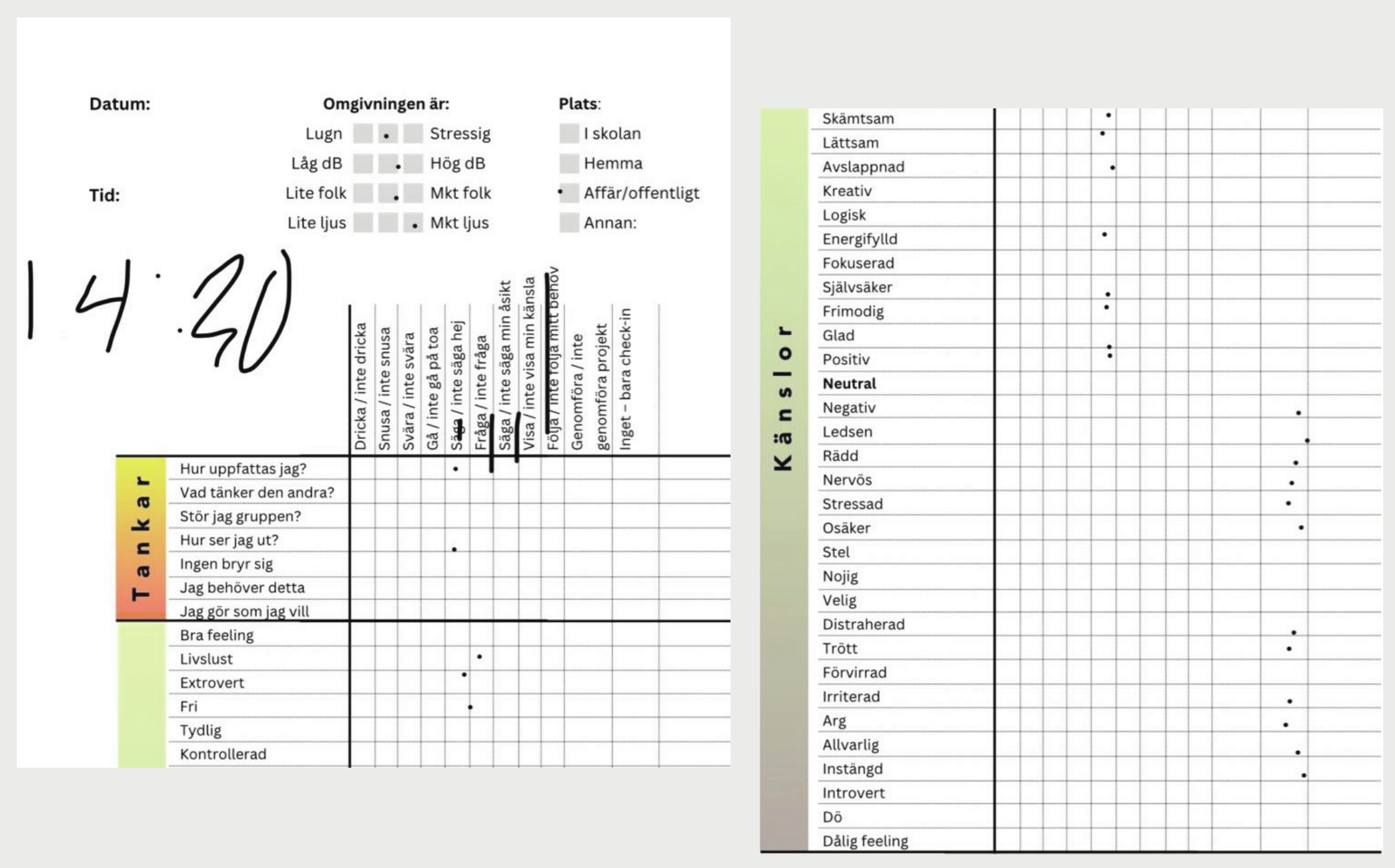}
  \caption[]{Excerpts from P5’s self-tracking setup. External contextual factors (date and time, location, and surroundings) are recorded at the top of the sheet. Below, the sheet is organized into two row categories: thoughts (orange) and emotions (green). Each column represents a contrasting pair, capturing an internal need alongside a corresponding masking response.}
  \Description{The figure shows a pair of images created by P5. The images show a screenshot divided in two to fit the page. In the screenshot, the top left reads date and time. Next to this are small cells organized by aspects of the physical context and place that can be annotated with 'ink'. For example, 'noise level' and 'at home'. Below, a larger table organizes thoughts and emotions with food, water and more. We include full details in supplemental materials as they do not fit here.}
  \label{fig:self-tracking-designs-p5}
\end{figure*}

All participants used their phone for tracking and tailored their approaches to contexts where masking felt salient. Although not prompted, each participant chose to track in an environment in which they had situated masking in their Phase I externalizations (P3: social situations; P5: school; P6: work). Their self-tracking approaches, shown in Figure \ref{fig:self-tracking-designs-p3-and-p6} and \ref{fig:self-tracking-designs-p5}, were shaped by distinct personal goals and varied in complexity and focus.
We describe each participant's approach in the following.

\textit{P3: Testing a hypothesis in event-based tracking.} 
P3 sought to test a hypothesis about masking less around neurodivergent friends and structured tracking around discrete social activities. 
He included a Likert scale to make it easy for himself: ``I wanted it to be a quick thing to fill in.'' 
Each time he engaged in a social activity, P3 used free text to log contextual details (date, activity description, duration, people present) and responded to six Likert scale items (1-5) covering energy levels, masking amount, discomfort, over-filtering, awareness of body language, and reacting as expected, plus one yes/no item (medication status) and space for additional comments. He completed 14 logs over one week.

\textit{P5: Capturing emotional complexity in a diagram sheet.} P5 aimed to capture the complexity of her emotions and identify patterns in how they related to masking. She logged hourly while at school, drawing on her reflections from Phase I: 
``I would need to conduct one [self-tracking log] every hour sometimes to find any average.''
Her framework included 64 items across six categories: location (4 items), surroundings (4 items), needs and wants (12 items), thoughts (7 items), and emotions (37 items). Using a pre-constructed diagram sheet, she marked which items were present each time she logged. She collected 34 logs over one week.

\textit{P6: Recognizing masking through minimal tracking.} P6 focused on simply recognizing when masking occurred, motivated by both challenges identifying their masking behaviors and difficulties maintaining routines: ``I just wanted to find out when it [masking] happened and to confirm that it had happened [...] I went down the \textit{Atomic Habits} route, perhaps, thinking that I would aim very low because I know that I find routines and strategies like this very difficult.'' P6 logged on workdays only. They used their existing digital work schedule as a tracking scaffold, marking an ``X'' on days when masking occurred and later adding free-text reflections about specific instances. They completed 5 logs over two weeks.

\subsubsection{What Participants Said}
While collecting and reflecting on data about a common topic (masking), participants' experiences with self-tracking were unique. Using participant quotes from exit interviews, we highlight findings related to their different approaches, existing routines, and personal perspectives on masking.

Detailed and simple approaches led to unique challenges and insights. P3 described a challenge in balancing his desire to ``include everything'' and ``add more questions'' with tendencies for overthinking. Though he attempted to make tracking easy for himself, he noted that he ``always [ended] up overthinking things and it might have been better with even fewer questions.'' In contrast, P6 found that having marked an X alone was often too vague to help them recall specific events, making it impossible to add later reflections in logs as intended. However, even without those reflections, this method had helped P6 to better understand their masking behaviors and brought attention to certain patterns. P5 found choosing between specific items limiting. Despite having tailored 64 items to reflect her masking behaviors, P5 explained that at times, no options matched her actual experiences. Attempting to interpret her results, P5 described her logs as mostly uninformative and overwhelming. However, she found that there were times when extensive tracking of specific emotions gave her insight into their complexity and how they are connected. She also valued the process of defining and using her own approach: ``I think it was still useful to at least try to do it every hour, even though I maybe did not do it every hour. But at least I got to try it and see.''

Available mental resources and existing routines affected self-tracking for all participants. P3 managed daily logging during the first week and attributed this consistency to being unemployed at the time. However, he chose to stop after the first week because he felt he was not gaining meaningful insights. P5, in contrast, expressed enthusiasm at the first week check-in about continuing tracking with planned modifications to simplify her framework. However, when she started an internship that same week, she was unable to implement these changes or continue tracking: ``It feels really stupid because I really wanted to do it [implement changes and continue self-tracking]. I started my internship last week and I was so super duper nervous. So I was sinking through the ground. I felt so bad and I thought about all the worst case scenarios.''

For P6, these challenges extended to long-standing difficulties with routines more broadly. They explained that finding mental resources and remembering to log in the evenings was challenging, particularly when tired from work. P6 contextualized this as a lifelong challenge and recounted previous unsuccessful attempts with psychiatric support to build routines using a journal and calendar. Drawing on this history, P6 attributed their challenges in self-tracking to personal difficulties rather than their approach: ``I've tried to stick to the method, but then it has just become what it has become. There hasn't been any change that has been justifiable as in `this doesn't work as well as this'. It's more been like, oh, I couldn't cope, oh, I forgot [to log], oh, it [self-tracking] didn't happen.'' At the same time, P6 reflected that staying connected with a group might have helped: ``Maybe having a group where you communicate regularly or with some regularity could be a good way to remind you that this [self-tracking] is something that is actually being done.''

Participants described diverse relationships with masking that became evident through their self-tracking. P3 sometimes viewed masking positively, which he incorporated into his tracking approach with a question about whether he wished he had masked more or less: ``Sometimes I see it [masking] as a kind of tool to be more social, especially with new people or in shops or whatever. You act polite and friendly in a way that feels fake, but sometimes I still want to do that, because it can lead to something more genuinely social in life, like actually getting to know people over time.'' In contrast, P5 reflected after tracking on the tension of masking confidence while struggling internally: ``It feels a bit wrong to give the impression that I am very confident and have everything together while I come home and cry. It makes it harder for people around me [\ldots] to understand me.'' For P6, self-tracking enabled new awareness of unwanted, habitual masking patterns: ``It's like biting your nails, if you've been doing it all your life, it's hard to stop.'' They described how self-tracking brought them closer to understanding these patterns: ``I used to think I must be a compulsive liar because I lie about meaningless things [when masking]. But they’re not meaningless. It’s about satisfying other people’s image of me instead of just being chaotic as I really am. So in that way, it has been rewarding to see the patterns I fall into when I mask.''

\section{Findings: Participant Reflections}
\label{sec:findings-participant-reflections}

Here, drawing on data from both study phases, we focus on participants' experiences and takeaways from our study as well as their reflections on masking, self-reflection, and personal data engagement. We organize these results into four themes that we formed during our analysis. We subtitle and illustrate each theme with representative quotations.

\subsection{Defining and Identifying Masking: ``It feels like I’ve been chasing shadows''}
Participants found it challenging to define and recognize masking, both conceptually and in relation to themselves. The ambiguous and frequently subconscious nature of masking contributed to this and made it difficult to track and reflect on.

\begin{quote}
``I am very interested in knowing how [to unmask]. It feels like my mask is glued to my face. It has probably grown attached.'' \textit{P6, externalization workshop (Phase I).}
\textit{Later, they said:}
``\textbf{It feels like I've been chasing shadows}. I catch myself afterward, usually long afterward, like several hours after I've done something [masking]. Oh, I did something. Then I write it down and try to think back on a fairly minor event. It's usually a very pointless, small and silly event.''
\textit{P6, exit interview (Phase II).}
\end{quote}

Difficulties defining when and how masking concretely occurs were raised by all participants and in both phases of the study. Masking by habit, they found it difficult to both distinguish what counts as masking and to understand when it happens. For instance, P6 theorized that they likely engaged in masking more frequently than they realized. Discussing the CAT-Q~\cite{hull2019development} in Phase I, P6 chose to share with the group that their result score had been ``low'' and said, ``Even when I take the test [CAT-Q], I have a low score. Because maybe I don't even know how much I have automated.'' 
Another example comes from P7, who reflected: ``It is difficult to draw the line or understand myself, when I am masking and what has blended together and become a part of me.'' A similar experience was shared by P4: ``I have gone through a lot with myself and realized that I have been masking my entire life [\ldots] In the beginning it was really difficult to understand the term [masking], it is still difficult today and especially when you relate it to yourself [\ldots] What am I really, and what is not me?''

Additionally, participants found that the ambiguous nature of masking contributed to challenges in defining and reflecting on such behaviors. For instance, P8 expressed a challenge in understanding masking as a concept in the externalization workshop and sought further clarification about how to interpret his CAT-Q scores in his follow-up interview. P3, who used a 1-5 Likert scale to track his ``amount of masking'' in Phase II, described in his exit interview how he had struggled with interpretation: did logging a ``1'' mean ``no masking'' or ``just a little''? He questioned, ``Can you ever be completely without masking?'' This ambiguity extended to other aspects of his tracking approach. On the last day of self-tracking, P3 removed the question ``Should I have masked more or less?'' which he disliked because he found it ambiguous, making it difficult to respond. He described this change as ``a relief not having to think about it [the question] anymore.'' Overall, P3 identified his central challenge in self-tracking as quantifying and reflecting on ambiguous experiences such as masking, noting that the vagueness sometimes made tracking feel ``too philosophical'' and unhelpful for deriving insights.

Both P3 and P8 highlighted the CAT-Q~\cite{hull2019development} as their preferred method to externalize masking in Phase I.
P3 also liked engaging with the drawing and writing prompts, but appreciated the questionnaire for its specific examples: ``Despite some frustrations over CAT-Q, I believe that it was most rewarding since there were more specific things to reflect on.''
For P8, this preference was also grounded in preferring to choose an option that fits him rather than ``to paint the significance of an emotion,'' due to the vagueness of such tasks. Elaborating, P8 described this as a general challenge:
``I found these exercises [drawing and writing prompts] very difficult to do because they were so open-ended [...] This is actually quite typical for people with Asperger's, too. But I often find it quite difficult to do tasks that don't have clear instructions. Or that don't have instructions at all, really. Because it was like, well, paint how this feels, and that feels like, yeah, okay.''

\subsection{The Problem(s) with Generalization: ``It depends''}
Participants addressed the complexity of context and emotion in relation to masking and self-tracking. They emphasized that their emotions, experiences, or behaviors, including masking, cannot be easily generalized.

\begin{quote}
``Sometimes I know the answer right away. But then it's usually a more specific question. But when it becomes more general, like, do I put a lot of energy into social interactions? What social interactions? Is it with my partner or is it with my teacher? Am I in class? Is it that I'm going to get on the bus? Is it that I'm going shopping? Is it that I bump into someone? So, then it becomes like this, social interactions. It's necessary for me to get on the bus sometimes, which has been really difficult. Where should I sit? What should I do? What should I say? Hi, should I say something? Should I take out my headphones? Should I take them out of my ears? Then it's difficult. But it's like the other day when I was going to pay for something in the store and I was really tired. Then I talked anyway. Then it wasn't difficult at all. So \textbf{it depends}.'' \textit{P5, externalization workshop (Phase I).}
\end{quote}

The tension between contextual nuance in experiences and the demand to generalize them in personal data practices was evident in the discussion of quantification and questionnaires in Phase I. When asked open-ended questions about their experiences with the CAT-Q~\cite{hull2019development}, questionnaires, and quantification in general, participants collectively began describing the negative emotions that these practices provoke in them. When discussing how ``social interactions'' cannot be generalized, participants noted how CAT-Q questions largely ask respondents to assess behaviors in relation to such generalized ``social interactions'' and ``social situations.'' Recognizing such abstractions as a common challenge for many neurodivergent individuals, P7 remarked: ``Perhaps they [assessment tools for ADHD and autism] should just ask one question at the end. Did you find it difficult to answer the questions? And that in itself is the answer to whether you have a diagnosis or not.''

Participants expressed that generalized questions cannot capture the situational subtleties in their experiences, as P6 described: ``I just feel like I'm answering questions that are not digging deep enough for me to feel like I'm answering for myself.'' Participants emphasized that it was important to provide authentic responses when producing any form of personal data, regardless of purpose. When asked to respond to generalized questions or to choose between generalized responses, they worried that their chosen response might not accurately reflect their intention. Similarly, participants described often feeling anxious about whether they were ``answering wrong'' when asked to give numerical ratings, as P7 explained: ``the fear is that the very square answer you have to give in the form of a number only is not interpreted in the way you intended.''

Participants described contextual assessment as central in deciding how to act, including masking. However, interpreting a situation required more than noticing physical surroundings: it was emotionally driven and often shifting from moment to moment. When self-tracking in Phase II, P3 distinguished logs based on how experiences ``felt,'' which brought attention to how he evaluates context in relation to masking:

\begin{quote}
``So I kind of distinguished [self-tracking logs] depending on whether it felt like another thing. Like yesterday, two friends and I went to a lot of different shops, but it felt like the same thing in each shop, and so I had it as one log. And then one day we were at a water park, and then I thought there was a difference between being at the relaxation center where everyone was like adult and fancy, so I also had to be like that, compared to in the slides when people just scream. But it also meant that there was room to overthink a lot.'' \textit{P3, exit interview (Phase II).}
\end{quote}

Several participants expressed how their emotional states and perceptions fluctuate and may rapidly change, making it difficult both to track their experiences accurately and to evaluate them in generalized terms. This variability could also make responses feel inauthentic. In the externalization workshop, P6 noted: ``I think I have such different answers all the time. It can be that I am completely against a question, and in the next one I am all for it.'' Similarly, P5 explained: ``I’m very emotional and stuff. So it varies a lot depending on how I feel that day and all that.'' 

\subsection{Overthinking Responses: ``Is this really how I feel?''}
\label{subsec:finding-overthinking-responses}
Participants expressed a common tendency for overthinking, which made personal data practices particularly challenging. They described the processes of producing personal data as time-consuming, stressful, and overwhelming.

\begin{quote}
``I'm usually very slow at things like this [questionnaires], no matter what they involve. Like [even] if it's my tenancy, questions about my opinions. And it says it takes one minute. And then I sit there for fifteen minutes and just think, \textbf{is this really how I feel?} Then I get annoyed that it says, on this test [the CAT-Q], of all tests, I thought it should say that it could take up to half an hour or something. But then it's five minutes. Five to ten.'' \textit{P3, externalization workshop (Phase I).}
\end{quote}

Participants explained how overthinking was frequently triggered when producing personal data. This often led to increased time consumption when, for instance, completing surveys or questionnaires. Due to the CAT-Q's~\cite{hull2019development} targeting of autistic audiences, P3 expressed dismay that this possibility is not addressed in its time estimate. Other participants agreed. For instance, P7 shared that he would sometimes ``get stuck on a question for one hour'' when completing surveys. P6, conversely, shared that they had developed a strategy to limit time spent on each question ``to actively avoid becoming stuck and overthinking.''

Participants described getting stuck due to overthinking as anxiety-provoking and uncomfortable. This experience stemmed from their desire to provide authentic responses combined with their fear of answering ``wrong'' or inauthentically. Attempting to answer accurately while finding it difficult to do so triggered uncertainty and a fear of making mistakes, leading to overthinking. Practices that demanded what participants found challenging --- such as responding to generalized questions, providing numeric ratings, or formalizing ambiguous concepts --- particularly triggered overthinking. Additional sources of difficulty included structurally complex or awkwardly phrased questions. For example, P5 described struggling with a double-negation item in the CAT-Q: ``I have to sit there and do maths and Swedish grammar. What is the purpose here?'' Internal consistency checks posed another challenge:

\begin{quote}
``For me, something like that [internal consistency checks] creates stress. I become insecure. What did I answer to the last question? What do they mean? What should I do?'' [P7: ``Or have you misinterpreted the last question?''] ``Yes, have I understood this correctly? What exactly am I answering? What is the intention? [\ldots] If I answer differently, am I making stuff up? Am I fake?'' \textit{P5, externalization workshop (Phase I).}
\end{quote}

Noticing repeated or rephrased questions prompted participants to second-guess their own responses and worry about why their answers did not always match across similar items. These checks often undermined participants' confidence in their responses and interpretations, but also in their identities, as demonstrated in the above quote. More broadly, when experiencing fluctuating emotions and giving dissimilar responses, participants questioned both the intention behind the repetition and whether the tool worked for them. P6 explained: ``It can feel like the test wants to put you in a box. Or it wants to identify, are you like that or are you not like that? For me, since I feel like I bounce around a lot from answer to answer, it feels like the test doesn't work for me. And that maybe makes me frustrated.''

\subsection{Lasting Impact from Participation: ``It feels like you have strength in numbers''}
\label{sec:lasting-impact}
Participants emphasized that one of the most rewarding aspects of their participation was the shared reflections during group discussions. Many described these conversations as leaving a lasting impact, particularly because opportunities to exchange perspectives with other neurodivergent individuals are rare.

\begin{quote}
``I really enjoyed breaking down CAT-Q. The first thing we did was probably my favorite part. Breaking down the test, so to speak. Because it feels like something you often take for granted, that the test is already unbreakable, that it cannot be criticized or analyzed critically. But the first thing that happened was that everyone started saying that the test [CAT-Q] made them stressed or that they felt bad. Or that they were completely lost and didn't even know what the questions were trying to ask. So I thought it was interesting to see. Because \textbf{it feels like you have strength in numbers} when everyone is sitting there with similar difficulties in life. And then you look at this thing that is meant for you. Then it feels like now we are the standard and now it is the test that is odd.''  
\textit{P6, follow-up interview (Phase I).}
\end{quote}

Across all follow-up interviews, participants consistently valued the opportunity to reflect and share with other neurodivergent individuals. Importantly, participants highlighted how seldom they find themselves in such contexts: ``You don’t experience these things too often, getting to talk to like-minded people'' (P4); ``I certainly haven't spent a lot of my life in contexts like this, where there are actually other people who have similar problems, or difficulties, or strengths, or skills, or however you want to frame it'' (P7); ``[It was most rewarding] to be in a room where the normal is that people are not neurotypical whatsoever'' (P6). In follow-up interviews, four participants (P3, P4, P5, P7) specifically stated that the workshop experience helped them ``not feel alone.'' When asked when he felt most engaged, P8 replied: ``When others were talking [...] I found it very interesting to listen.''

Participants also described ways the workshop experience shaped their outlooks and behaviors. For P7, it sparked reflections on the value of unmasking: ``Even though I understand the importance and value of living a life where my exterior reflects my interior, I am so beaten down and devastated by the feeling that I can't go to a job interview and be myself. I can't go to my job and be myself. I can't get on the bus and be myself. [Supporting unmasking is about] highlighting the advantages in a way that overpowers that feeling.''

Following the workshop, P3 shared in his interview that he felt encouraged both to challenge himself to unmask and to discuss masking more openly with friends. P6 noted that having an outlet to share frustrations and strategies could prevent negative feelings from accumulating while adding nuance to one's own perceptions. During the reflection and ideation stage in the workshop, they shared: ``Personally, I think that everyone has very individual interpretations of their difficulties, even though the difficulties may be the same [\ldots] And that means you might start to reformulate your own interpretation as well, nuance your own interpretation. It can affect how you live your life afterward, and that's perhaps the most important thing.''

P8 described the self-recognition he had gained from participation as both anxiety-provoking and rewarding. In his follow-up interview, P8 shared that he had no prior knowledge of the CAT-Q~\cite{hull2019development} and said, ``I was extremely surprised and actually felt somewhat, well, a bit anxious about how high I felt I scored on it.'' He then added: ``I think that throughout my life I've actually wanted to distance myself a little from having Asperger's. [...] And I probably still feel that way a little, that it would be nice if it weren't like that. And then it's a little hard to get that as a result [on the CAT-Q]. Just like this, yes, you probably do this [mask] a lot more than you thought.'' At the same time, P8 found the externalization exercises in the workshop interesting and rewarding, particularly in relating to others' experiences with masking: ``I found it very rewarding to hear about other people's experiences [...] I've always thought personally before this that I didn't mask that much. But there was a lot that came up that I could really relate to.''\\

\section{Discussion}
We discuss our findings and the challenges we identified, and outline opportunities for responding to them through further empirical work, through exploring design opportunities, and through advancing theory in the intersection of neurodiversity and personal informatics.

Based on a synthesis of our findings, we first introduce a working model of emotion in personal informatics comprised of three emotional dimensions. 
Next, we describe and discuss each of these dimensions, including the challenges they pose, consequences of them, and design opportunities towards addressing them.
Further, we outline directions for future work to build on our findings and insights.

\subsection{Introducing a Working Model of Emotional Dimensions in Personal Informatics }
\label{sec:hypothesis}
Our findings demonstrate ways in which creating and reflecting on personal data can be emotionally demanding and thus exert emotional labor.
We note how participants describe such challenges in relation to self-tracking for self-reflection, but also more broadly in producing personal data.
To better understand the observed phenomena and respond to these challenges, we next seek to disentangle participants' experiences through three perspectives.
First, we notice how participants describe an \textit{emotional weight in responding}. 
Second, we notice how participants describe a \textit{discomfort from emotional self-reflection}. 
Third, we notice how participants describe an \textit{emotional burden from obligations} in self-tracking.
We present a brief overview of these perspectives as three emotional dimensions that shaped participants’ experiences with creating and reflecting on personal data.

\textbf{Emotional weight} refers to immediate negative emotions that arise during the act of externalization or self-tracking and are tied to overthinking and uncertainty. Participants described fear of misunderstanding a question, making mistakes, or appearing inauthentic in responses. They experienced uncertainty and stress when faced, for example, with unclear terminology, lack of context, or internal consistency checks.
    
\textbf{Emotional self-reflection} describes discomfort provoked by the insights that self-tracking can surface. For example, one participant (P8) struggled with a high score on the CAT-Q~\cite{hull2019development} as it forced the recognition of an autistic identity he wished to distance himself from. While challenging, such reflections also signal the potential of self-tracking to support meaningful personal growth.

\textbf{Emotional burden} captures negative emotions associated with expectations around longer-term tracking and is tied to a sense of obligation. It extends beyond the act of tracking and may also occur in, for instance, anticipating future tracking which one expects to cause an emotional weight. One participant (P3) described relief when reducing their expectations for self-tracking by removing a question, while others (P5, P6) expressed how they found self-tracking adherence mentally demanding.

We introduce these dimensions as a working model or hypothetical theory to be tested and validated in future work. 
We discuss opportunities for expanding the scope and validating this model in later sections.
In the following, we describe and discuss each of the three dimensions.

\subsection{Emotional Weight: When Responding Leads to Overthinking}
\label{sec:emotional-weight}
We notice the immediate stress and overthinking that can occur during the act of externalizing or tracking personal data. 
Participants feared misunderstanding questions, choosing the ``wrong'' option, or appearing inauthentic. These reactions were amplified by vague or complex terminology and a lack of contextual scaffolding. We also identified internal consistency checks (i.e., repeated or rephrased items intended to validate responses) as imposing emotional weight. 

Internal consistency checks are commonly applied in questionnaires to validate responses~\cite{broomell2014parameter}. However, in contrast to this purpose, their inclusion has been suggested to reduce response validity. Li et al.~\cite{li2022more} used mouse- and eye-tracking to examine how respondents react to such checks and found that they became less attentive, simplified the task, and processed less information. Interestingly, our findings show the opposite: for our participants, noticing internal consistency checks caused them to overthink, to make the task more complex, and to process more information. The repetition provoked participants to overanalyze both questions and responses and, if their answers to similar questions differed, to question themselves. 

Recommendations for accessible survey instruments~\cite[e.g.,][]{nicolaidis2020creating} emphasize the importance of clarity in the description of questionnaire items. However, to our knowledge, overthinking triggered by internal consistency checks in neurodivergent contexts remains unexamined. Our findings suggest that this routine methodological practice can disproportionately burden autistic and ADHD individuals and limit their representation in questionnaire results (a form of epistemic injustice), calling for further exploration in future work.

Participants largely described overthinking as inevitable and leading to excessive time consumption in personal data practices. Thus, short completion-time estimates in instruments like the CAT-Q~\cite{hull2019development} introduced pressure at the outset and further increased emotional weight. We interpret such estimates as normative expectations that do not accommodate diverse cognitive processing patterns. Technologies often encode a ``status quo'' that privileges dominant assumptions~\cite{birhane2021algorithmic}. In this example, the ``status quo'' assumes that respondents can complete items quickly and decisively. This assumption persists even though the CAT-Q~\cite{hull2019development} is specifically designed for autistic individuals. Participants in our study described their actual engagement as requiring significantly more time and reflection than standard completion estimates suggest. Read through a Crip HCI lens, these defaults risk making standard instruments feel out of reach at the very point of entry. The mismatch between tool design assumptions and how participants actually engaged risks transforming a supportive tool into a source of additional pressure.

\subsection{Emotional Self-Reflection: When Self-Discovery is Difficult}
\label{sec:emotional-self-reflection}
We notice the discomfort that insights from self-tracking might provoke. While difficult, such moments are important in unmasking and other self-reflective processes.

Emotional self-reflection is not unique to neurodivergent contexts: Thudt et al.~\cite{thudt2018selfreflection} describe how participants reported ``disappointment'' and ``regret'' when reflecting on past actions, but still found these moments ``useful'' for self-understanding. For ADHD and autistic individuals, however, such experiences are often more mentally demanding~\cite{lin2022social,soler-gutierrez2023evidence}, and masking in particular is linked to fatigue and long-term mental health consequences~\cite{russo2018costs}. Because reflection is both inevitable and potentially beneficial, the challenge is not to suppress difficult emotions but to create conditions where individuals can engage with them safely and on their own terms. 
To be clear, this does not call for attempts to mitigate reflection itself. Rather, it calls for methods that support and empower people in self-tracking, \textit{even} when insights might be difficult to handle. 

However, designing generalized self-tracking support that helps navigate and manage emotions arising from self-tracking insights might be challenging. 
This holds especially when considering the complex emotional variability and context-dependencies experienced by ADHD and autistic individuals.
Our participants’ reflections were entangled with contexts that are difficult to formalize or generalize. In Phase I, they all agreed that ``it depends,'' and in Phase II, P3 noted that ``context'' exceeded physical surroundings, entailing shifting internal states and social dynamics. Throughout, participants expressed that it was difficult to define and identify masking in the first place and found it challenging to track such behaviors. Relationships to masking also varied: as seen in Phase II, P3 framed masking as ``a tool to be more sociable,'' whereas P6 had worried, ``I must be a compulsive liar because I lie about meaningless things.'' These diverse framings shaped what counted as insight, harm, or progress for each person.

As the experiences of our participants demonstrate, masking behaviors are ambiguous, difficult to define and identify, and challenging to make sense of in self-tracking. Additionally, neurodivergent experiences are complex and diverse, and engaging in data practices to understand them requires mental resources, which can be challenging when those resources fluctuate. Ai et al. asked about the ``optimal ways'' to measure impression management (masking) in ways that respect its dynamic, iterative, and context-dependent nature~\cite{ai2022reconsidering}. Our findings suggest that, for self-reflective purposes, effectiveness does not lie in universal `optimal' methods but in approaches that can adapt to changing contexts, preferences, and needs. Abilities and preferences for when, where, and how to track vary both between individuals and over time. Dynamic factors such as energy, emotional state, social context, and relationships with masking continually reshape what approaches are both feasible and valuable. 

\subsection{Emotional Burden: When Expectations Turn Into Pressure}
\label{sec:emotional-burden}
We notice the pressure that accumulates around self-tracking: felt in anticipation of future logging, in reflections on past efforts, and in the sense of `duty' to keep up.
These burdens persist across time and are often intertwined with perfectionism, difficulties maintaining routines, and self-criticism. 

In Phase II, participants described how self-tracking adherence required available mental resources and expressed ``relief'' when removing items, noting how it eased the mental load. Similar shifts toward self-compassion are echoed in prior work where a participant reported ``relief'' after loosening expectations during self-tracking~\cite{thudt2018selfreflection}.
This pattern contrasts with prevailing accounts of personal informatics. Epstein et al.’s work characterizes common difficulties as \textit{forgetting} to wear tracking devices or dropping routines over time~\cite{epstein2016reconsidering}. Choe et al.~\cite{choe2014understanding} note that tracking ``too many things'' can be manageable early on (when motivation is high) but often leads to tracking fatigue later. In our study, however, such a burden was present \textit{from the start}: the immediate emotional weight of overthinking added to the ongoing pressure of balancing mental resources with routines.

Self-tracking is often framed as a pipeline: collect, reflect, then change. Our findings suggest that, when tracking complex and emotionally driven behaviors like masking, reflection itself can be a sufficient and valuable outcome. While unmasking is sometimes driven by a desire to change behavior, it is inherently grounded in self-reflection and self-understanding~\cite{pearson2021conceptual}. Our participants reported gaining new insights and value, and even feeling encouraged toward future change, simply through articulating, externalizing and reflecting on experiences. Prior work similarly shows how making and engaging with personal representations can cultivate self-understanding and self-compassion as worthwhile ends~\cite{thudt2018selfreflection}. Participants who took part in our externalization workshop echoed this sentiment.

Many self-tracking systems in HCI center on behavior change, emphasizing explicit goals and routine maintenance~\cite{lee2017selfexperimentation}. However, our findings suggest that sustained, high-commitment tracking is not always necessary for meaningful insight. For our participants, even a one-time workshop involving externalization and reflection on masking was sufficient to spark new self-understanding. Additionally, one participant (P6) used a simple tracking method and occasionally forgot to log, yet still gained valuable insights about their masking behaviors. These examples suggest that when the experience being tracked is complex and ambiguous, as masking is, simple methods can serve as a counterbalance to that complexity. As seen in our study, lightweight personal data practices hold potential to reduce cognitive load while still yielding meaningful insights.

\subsection{Peer Support and Emotional Dimensions}
\label{sec:peer-support}
Drawing on the emotional dimensions and participants' emphasis on the value of peer support, we discuss the potential of integrating these aspects into managing self-tracking.
Peer contexts can offset emotional weight by normalizing challenges and providing support, deepen emotional self-reflection by offering validation and new perspectives, and reduce emotional burdens through shared accountability.

\textbf{Peer support and emotional weight.} When participants individually completed the CAT-Q~\cite{hull2019development}, they experienced overthinking and stress. However, in the Phase I workshop, collectively analyzing the same questionnaire and recognizing shared frustrations became one of the most valued aspects of participation. Participants expressed how, for instance, generalized questions and internal consistency checks had made them feel anxious. In a peer context, these frustrations were validating rather than isolating. Of six participants, four specifically expressed appreciation for ``not feeling alone'' when sharing masking experiences in a context where \textit{they} decided the defaults and defined the norms. This collective recognition transformed what had been an individual weight into a shared, normalized experience.

\textbf{Peer support and emotional self-reflection.} Peer support also eased the mental load in navigating emotional self-reflection around ambiguous, complex masking experiences. This aligns with earlier work highlighting the benefits of shared reflection~\cite{thudt2018selfreflection}, but our findings suggest that support from neurodivergent peers specifically brings both validation and new perspectives that contribute meaningfully to self-reflective practices like unmasking. For example, when one participant's (P8) high CAT-Q score provoked anxiety about recognizing an autistic identity he wished to distance himself from, relating to peers' masking experiences through workshop externalizations provided comfort. Participants expressed how having an outlet to share perspectives could affect their outlooks and behaviors. Even though we did not ask participants directly about the value of peer sharing, every single participant highlighted their appreciation for this in follow-up interviews.

\textbf{Peer support and emotional burden.} While we did not directly test peer-supported tracking approaches in Phase II, our findings suggest this as a promising direction. One participant (P6) shared that previous clinical journaling interventions to establish routines had been unsuccessful, contributing to an outlook that no self-tracking method could overcome their difficulties with routines. Yet they expressed interest in trying a peer-supported approach, noting that regular group communication might provide helpful reminders. Accountability mechanisms such as body doubling and accountability partnering are used in mental health domains to support task completion and routine maintenance~\cite{arnold2025individual,eagle2024it}. Given that emotional burden --- the pressure and obligation around maintaining tracking --- was a consistent challenge for participants, integrating such peer-based accountability mechanisms into self-tracking practices warrants further exploration.

\subsection{Future Work}
\label{sec:future-work}
Our work identifies new phenomena in relation to self-tracking and personal informatics more broadly.
Based on our findings, we begin to specify where emotional demands emerge in self-tracking practices and how design assumptions around stability, generalizability, and individual reflection contribute to them.
Considering our work identifies phenomena not previously described, our model and its hypotheses are necessarily limited in scope, in both breadth (future work may identify additional dimensions) and in depth (future studies may further elaborate how these dimensions unfold across contexts and over time).

We encourage future work to examine and extend our hypotheses with larger and more diverse participant groups, using both qualitative and quantitative approaches.
Such work could refine our understanding of for whom, when, and under what conditions emotional challenges arise, as well as explore design strategies in response to these challenges.
This way, our work might pave the way for engaging more deeply with emotional and cognitive diversity in personal informatics and thus mark a new research direction for the field and for HCI more broadly.

Building on our findings, we suggest three principles for self-tracking design in neurodivergent contexts. First, treat \textit{reflection as a sufficient goal}: participants reported value and encouragement toward future change simply from reflecting and discussing in a workshop, even without achieving concrete outcomes. Second, embrace \textit{simple methods and short-term goals}: smaller commitments reduce anticipatory pressure and make it easier to adapt when energy varies. Third, \textit{prioritize adaptability}: external factors and internal states constantly influence preferences and abilities in self-tracking, requiring flexible approaches rather than rigid routines. These principles echo the articulation by Spiel et al.~\cite{spiel2022adhd} of ``taking up crip time,'' which we recognize in both our own research processes and participants’ self-tracking. As such, we note the need to carefully balance flexibility with the need for clear, non-ambiguous instructions, and with the risk that constant adjustment itself becomes burdensome.

Moreover, we note our surprise at the extent of criticism voiced about the CAT-Q~\cite{hull2019development} and questionnaires more broadly.
Though we did not set out to study questionnaire methodology and cannot claim expertise in this area, we think the issues raised by participants are important to consider in future work.
For example, reading the transcripts with this in mind, we wonder whether and how different user interfaces might alleviate (or exacerbate) these experiences of questionnaires.
Although recent work has raised similar concerns about questionnaires~\cite[e.g.,][]{uglik-marucha2026fit}, future work might explore design opportunities for addressing these issues.

Our findings suggest that engaging with peers as part of self-tracking can help combat and potentially overcome challenges experienced in creating and reflecting on
personal data.
Thus, future work might beneficially explore the deliberate use of peer support in self-tracking tools and systems.
Promising opportunities include exploring how self-tracking might be integrated into structured group reflection, how sharing features in existing self-tracking technologies might be consciously designed from the perspective of peer support, and how mechanisms inspired by body doubling and accountability partnering might play a more central role. Rather than positioning self-tracking as an individual practice that must be navigated alone, situating it in community contexts can help offset emotional weight, deepen self-reflection, and reduce burdens to transform difficult processes into connection and growth.

Additionally, our findings and principles may extend to self-tracking and self-reflection practices related to mental health in general.
This echoes self-tracking work with bipolar individuals, in which authors highlight the need for personalized tracking and extend this to technology-based monitoring of mental health more broadly~\cite{murnane2016selfmonitoring}. 
In this domain, we think that principles like focusing on reflection as well as benefits from peer support apply meaningfully. We find it particularly relevant to extend our findings to other invisible disabilities, where navigating authenticity and identity concealment can be a similar challenge. In these contexts, we theorize that peer support can be especially valuable when carefully facilitated, as individuals are prompted to openly reflect with others on emotional and highly personal experiences which they tend to conceal.

Finally, we think that these opportunities can benefit communities and individuals beyond neurodivergent contexts. Particularly, such benefits could extend to other marginalized groups.
Similar to masking, concealment of sexual or gender identity is consistently linked to negative impacts on mental health. Kiekens and Mereish~\cite{kiekens_association_2022} highlight this issue and examine peer support in this context. They found that peer support did not significantly affect the relationship between daily identity concealment and negative affect. However, they note an explanation for this: they measured general emotional support (e.g., ``I can count on my friends when things go wrong'') rather than identity-specific support, and did not assess whether peers shared the marginalized identity. This distinction is crucial for what we propose: intentionally facilitated peer interactions around shared identity-related challenges.

\subsection{Strengths and Limitations}
\label{sec:strengths-limitations}
As neurodivergent researchers, we approached this work from our lived experiences within the communities we study. We view this positionality as both an analytical resource and a potential limitation. Our insider perspective shaped what we recognized as significant findings and how we interpreted participants' accounts. While this facilitated nuanced understanding of complex experiences like masking, it also introduced interpretive biases that we cannot fully disentangle from the analysis.
Yet, because of this, we could deeply understand and empathize with participants' accounts of, for example, overthinking and masking behaviors. No doubt, in large part because we navigate these same challenges. We participated alongside participants in Phase I externalization exercises to establish shared ground and mutual vulnerability, though we excluded our own data from analysis to avoid conflating researcher perspectives with participant experiences and to center participant voices over our own preexisting knowledge and expectations.

Our participants were recruited from our personal networks, resulting in a relatively narrow and specific group of neurodivergent individuals. This convenience sample primarily reflects young adults with ADHD and/or autism who were comfortable with verbal and visual expression, and who shared social proximity to the researchers. As such, the study does not aim to represent the full diversity of neurodivergent experiences, but rather to provide exploratory, situated insights into masking and self-tracking. We see this as both a limitation in terms of representativeness and a strength in terms of the depth and relatability of the accounts shared.

Small sample sizes are well-established in qualitative research, particularly with marginalized populations and when exploring sensitive phenomena~\cite{ting2023less}. 
We decided to not invite additional participants when the sample decreased after the first phase. 
We had already built trust and mutual understanding with the original group that we did not wish to jeopardize. 
Noting the strong link between what participants did in the two phases confirms our decision. 
While our small sample reduces the potential for generalizability or recounting insights as recurring tendencies, this was not our goal. 
Aligning with recommendations for qualitative research focusing on marginalized groups, we prioritized highlighting participants' voices and lived experience over generalizability. 
While the themes we identify will benefit from further investigation, our study contributes early qualitative insight into important phenomena.
Thus, our work can inform future work on self-tracking in neurodivergent contexts and potentially, in contexts with other marginalized groups.

\section{Conclusion}
We examined how neurodivergent individuals experience creating and reflecting on personal data about ambiguous complex behaviors, focusing on masking among autistic and ADHD individuals. 
Our findings show that personal data practices, which often presume stability or generalizability, can place substantial interpretive and emotional demands on individuals. 
Our participants’ experiences point to three emotional dimensions that shape engagement with personal data, which we present as a working model and conceptual lens for understanding emotional challenges in personal informatics. 
We also found that some of these challenges often became manageable through collective sense-making in peer-supported contexts.
Based on these findings, we argue that personal informatics should engage more with social and emotional aspects, as well as with cognitive and emotional diversity. 
Designing more accessible self-tracking means attending more 
closely to the situated nature of self-tracking and more carefully considering how systems impact the people that use them.
Our work invites future studies to examine how emotional dimensions shape personal data practices and how peer-supported approaches may transform these practices, both in neurodivergent contexts and in HCI more broadly.

\begin{acks}
We would like to thank the people that took part in our study. We are grateful that they devoted their time and energy. 
Additionally, we would like to thank our colleagues, collaborators, and not least, the anonymous reviewers. Without doubt, they helped improve our paper.
\end{acks}

\bibliographystyle{ACM-Reference-Format}
\bibliography{references}

\end{document}